\RequirePackage{fix-cm}

\documentclass[twocolumn,epjc3]{svjour3}  

\smartqed  

\RequirePackage{graphicx}
\RequirePackage[numbers,sort&compress]{natbib}
\RequirePackage[colorlinks,citecolor=blue,urlcolor=blue,linkcolor=blue]{hyperref}
\RequirePackage{siunitx}
\usepackage{amsmath}
\usepackage{soul}
\usepackage{float}
\usepackage{graphicx, caption}
\usepackage{subcaption}
\usepackage{comment}
\usepackage{accents}
\usepackage{amsmath}
\usepackage{tikz}
\usepackage[compat=1.1.0]{tikz-feynman}
\usetikzlibrary{arrows.meta, decorations.markings, decorations.pathmorphing}
\usepackage{amssymb}
\usepackage{siunitx}

\usepackage{ragged2e}

\makeatletter

\patchcmd{\@maketitle}
  {\noindent\ignorespaces\@author}
  {\begin{center}\justifying\ignorespaces\@author\end{center}}
  {}{}
\makeatother

\usepackage[T1]{fontenc}
\usepackage[utf8]{inputenc}
\usepackage{lmodern}
\usetikzlibrary{arrows.meta, decorations.markings, decorations.pathmorphing}
\newlength{\dhatheight}
\newcommand{\doublehat}[1]{%
    \settoheight{\dhatheight}{\ensuremath{\hat{#1}}}%
    \addtolength{\dhatheight}{-0.35ex}%
    \hat{\vphantom{\rule{1pt}{\dhatheight}}%
    \smash{\hat{#1}}}}

\journalname{Eur. Phys. J. C}

\begin{document}

\onecolumn

\title{Search for two-neutrino double electron capture in $^{36}$Ar with the DarkSide-50 detector}

\subtitle{DarkSide-50 collaboration\thanksref{email_jpc}}

\author{
Paolo Agnes\thanksref{aff1,aff2} \and
Ivone F.M. Albuquerque\thanksref{aff3} \and
Thomas Alexander\thanksref{aff4} \and
Andrew Knight Alton\thanksref{aff5} \and
Maximo Ave Pernas\thanksref{aff1} \and
Henning Olling Back\thanksref{aff4}\thanks{Present address: Savannah River National Laboratory, Jackson, SC 29831, United States} \and
Giovanni Batignani\thanksref{aff6,aff7} \and
Walter Marcello Bonivento\thanksref{aff8} \and
Bianca Bottino\thanksref{aff9,aff10} \and
Severino Bussino\thanksref{aff11,aff12} \and
Matteo Cadeddu\thanksref{aff8} \and
Mariano Cadoni\thanksref{aff8} \and
Alessio Caminata\thanksref{aff9} \and
Nicola Canci\thanksref{aff13} \and
Mauro Caravati\thanksref{aff8} \and
Nicola Cargioli\thanksref{aff8} \and
Marco Carlini\thanksref{aff2} \and
Sergey Chashin\thanksref{aff39} \and
Alexander Chepurnov\thanksref{aff39} \and
Stefano Davini\thanksref{aff10} \and
Sandro De Cecco\thanksref{aff14,aff15} \and
Alexander Derbin\thanksref{aff39} \and
Daniel D\'iaz Mairena\thanksref{aff16} \and
Francesca Dordei\thanksref{aff8} \and
Giuliana Fiorillo\thanksref{aff17,aff13} \and
Davide Franco\thanksref{aff18} \and
Federico Gabriele\thanksref{aff8} \and
Cristiano Galbiati\thanksref{aff19} \and
Graham Kurt Giovanetti\thanksref{aff20} \and
Maxim Gromov\thanksref{aff39} \and
Marisa Gulino\thanksref{aff21,aff22} \and
Brianne Rae Hackett\thanksref{aff4} \and
Fabrice Hubaut\thanksref{aff23} \and
Andrea Ianni\thanksref{aff19} \and
Valerio Ippolito\thanksref{aff15} \and
Feodor Karpeshin\thanksref{aff39} \and
Denis Korablev\thanksref{aff39} \and
George Korga\thanksref{aff24} \and
Michael Kuss\thanksref{aff6} \and
Marco La Commara\thanksref{aff25,aff13} \and
Michela Lai\thanksref{aff26} \and
Marcello Lissia\thanksref{aff8} \and
Olga Lychagina\thanksref{aff39} \and
Igor Machulin\thanksref{aff39} \and
Stefano Maria Mari\thanksref{aff11,aff12} \and
Jelena Maricic\thanksref{aff27} \and
Radovan Milincic\thanksref{aff27} \and
Matteo Morrocchi\thanksref{aff6} \and
Valentina Muratova\thanksref{aff39} \and
Paolo Musico\thanksref{aff10} \and
Marco Pallavicini\thanksref{aff9,aff10} \and
Luciano Pandola\thanksref{aff21} \and
Emilija Pantic\thanksref{aff28} \and
Eugenio Paoloni\thanksref{aff6,aff7} \and
Krzysztof Pelczar\thanksref{aff29} \and
Vicente Pesudo\thanksref{aff16} \and
Andrea Pocar\thanksref{aff30} \and
Stephen Pordes\thanksref{aff31} \and
Pascal Pralavorio\thanksref{aff23} \and
Marco Razeti\thanksref{aff8} \and
Andrew Lee Renshaw\thanksref{aff32} \and
Marco Rescigno\thanksref{aff15} \and
Davide Sablone\thanksref{aff33} \and
Oleg Samoylov\thanksref{aff39} \and
Simone Sanfilippo\thanksref{aff20} \and
Roberto Santorelli\thanksref{aff16} \and
Claudio Savarese\thanksref{aff34} \and
Dmitriy Semenov\thanksref{aff39} \and
Andrei Sheshukov\thanksref{aff39} \and
Mikhail Skorokhvatov\thanksref{aff39} \and
Oleg Smirnov\thanksref{aff39} \and
Albert Sotnikov\thanksref{aff39} \and
Simone Stracka\thanksref{aff6} \and
Yury Suvorov\thanksref{aff17,aff13} \and
Roberto Tartaglia\thanksref{aff2} \and
Gemma Testera\thanksref{aff10} \and
Evgeniy Unzhakov\thanksref{aff39} \and
Alina Vishneva\thanksref{aff39} \and
Bruce Vogelaar\thanksref{aff31} \and
Masayuki Wada\thanksref{aff35} \and
Yi Wang\thanksref{aff36,aff37} \and
Shawn Westerdale\thanksref{aff38} \and
Marcin Marian Wojcik\thanksref{aff29} \and
Changgen Yang\thanksref{aff36,aff37} \and
Grzegorz Zuzel\thanksref{aff29}
}

\institute{
Gran Sasso Science Institute, L'Aquila, 67100, Italy \label{aff1} \and
INFN Laboratori Nazionali del Gran Sasso, Assergi (AQ), 67100, Italy \label{aff2} \and
Instituto de F\'isica, Universidade de S\~ao Paulo, S\~ao Paulo, 05508-090, Brazil \label{aff3} \and
Pacific Northwest National Laboratory, Richland, WA 99352, USA \label{aff4} \and
Physics Department, Augustana University, Sioux Falls, SD 57197, USA \label{aff5} \and
INFN Pisa, Pisa, 56127, Italy \label{aff6} \and
Physics Department, Universit\`a degli Studi di Pisa, Pisa, 56127, Italy \label{aff7} \and
INFN Cagliari, Cagliari, 09042, Italy \label{aff8} \and
Physics Department, Universit\`a degli Studi di Genova, Genova, 16146, Italy \label{aff9} \and
INFN Genova, Genova, 16146, Italy \label{aff10} \and
INFN Roma Tre, Roma, 00146, Italy \label{aff11} \and
Mathematics and Physics Department, Universit\`a degli Studi Roma Tre, Roma, 00146, Italy \label{aff12} \and
INFN Napoli, Napoli, 80126, Italy \label{aff13} \and
Physics Department, Sapienza Universit\`a di Roma, Roma, 00185, Italy \label{aff14} \and
INFN Sezione di Roma, Roma, 00185, Italy \label{aff15} \and
CIEMAT, Centro de Investigaciones Energ\'eticas, Medioambientales y Tecnol\'ogicas, Madrid, 28040, Spain \label{aff16} \and
Physics Department, Universit\`a degli Studi ``Federico II'' di Napoli, Napoli, 80126, Italy \label{aff17} \and
APC, Universit\'e de Paris, CNRS, Astroparticule et Cosmologie, Paris, F-75013, France \label{aff18} \and
Physics Department, Princeton University, Princeton, NJ 08544, USA \label{aff19} \and
Williams College, Department of Physics and Astronomy, Williamstown, MA 01267, USA \label{aff20} \and
INFN Laboratori Nazionali del Sud, Catania, 95123, Italy \label{aff21} \and
Engineering and Architecture Department, Universit\`a di Enna Kore, Enna, 94100, Italy \label{aff22} \and
Centre de Physique des Particules de Marseille, Aix Marseille Univ, CNRS/IN2P3, CPPM, Marseille, France \label{aff23} \and
Department of Physics, University of Oxford, Oxford, OX1 3RH, UK \label{aff24} \and
Pharmacy Department, Universit\`a degli Studi ``Federico II'' di Napoli, Napoli, 80131, Italy \label{aff25} \and
Department of Physics, Engineering Physics and Astronomy, Queen's University, Kingston, ON K7L 3N6, Canada \label{aff26} \and
Department of Physics and Astronomy, University of Hawai'i, Honolulu, HI 96822, USA \label{aff27} \and
Department of Physics, University of California Davis, Davis, CA 95616, USA \label{aff28} \and
M. Smoluchowski Institute of Physics, Jagiellonian University, Krakow, 30-348, Poland \label{aff29} \and
Amherst Center for Fundamental Interactions and Physics Department, University of Massachusetts, Amherst, MA 01003, USA \label{aff30} \and
Virginia Tech, Blacksburg, VA 24061, USA \label{aff31} \and
Department of Physics, University of Houston, Houston, TX 77204, USA \label{aff32} \and
Physics Department, Temple University, Philadelphia, PA 19122, USA \label{aff33} \and
Center for Experimental Nuclear Physics and Astrophysics, and Department of Physics, University of Washington, Seattle, WA 98195, USA \label{aff34} \and
AstroCeNT, Nicolaus Copernicus Astronomical Center of the Polish Academy of Sciences, Warsaw, 00-614, Poland \label{aff35} \and
Institute of High Energy Physics, Beijing, 100049, China \label{aff36} \and
University of Chinese Academy of Sciences, Beijing, 100049, China \label{aff37} \and
Department of Physics and Astronomy, University of California, Riverside, CA 92507, USA \label{aff38} \and
ORCID:0000-0002-9159-6801, ORCID:0000-0002-1767-1754, ORCID:0000-0002-4351-2255, ORCID:0000-0003-2869-2363, ORCID:0000-0003-2161-1153, ORCID:0000-0002-4222-9650, ORCID:0009-0009-0770-8830, ORCID:0000-0002-0597-2234, ORCID:0000-0001-5532-7711, ORCID:0000-0003-2141-8230, ORCID:0009-0005-0286-0156, ORCID:0000-0001-5128-9279, ORCID:0000-0002-5527-4880, ORCID:0000-0002-5238-0442, ORCID:0000-0001-8371-5949, ORCID:0000-0003-2952-6412, ORCID:0000-0002-2624-9416 \label{aff39}
}

\thankstext{email_jpc}{e-mail: \href{mailto:ds-ed@lists.infn.it}{ds-ed@lists.infn.it}}

\date{Received: date / Accepted: date}

\maketitle

\twocolumn

\thankstext{e1}{e-mail: olia.lychagina@jinr.ru}

\begin{abstract}
Two-neutrino double electron capture is a rare nuclear decay where two electrons are simultaneously captured from the atomic shells and two neutrinos are emitted. We report on the first search for two-neutrino double electron capture in the \textit{KK}- and \textit{KL}-shells of $^{36}$Ar using the low-radioactivity liquid argon target from underground sources in the \mbox{DarkSide-50} experiment. No statistically significant excess was observed with approximately 12 ton-day exposure of underground argon (UAr) and, taking into account the $^{36}$Ar isotopic abundance in UAr (0.007\%), we set a limit on the half-life of the two-electron capture process in $^{36}$Ar of $T_{1/2}>\SI{9.2E19}{yr}$ at \mbox{90\% C.L.}
The sensitivity of the DarkSide-20k experiment, which will become operational in the next few years, was also evaluated and is expected to increase by a factor $\sim$100 with 10 years of expected operation and assuming the same $^{36}$Ar abundance as in the \mbox{DarkSide-50} underground argon target.
\end{abstract}
\section{Introduction}
\label{intro}
Double electron capture (2EC) is a class of rare second order weak-interaction processes with predicted half-lives exceeding the age of the Universe by more than 18 orders of magnitude~\cite{Winter_theory}. Two distinct reaction types may be identified in this class: those with (2EC2$\nu$) and without (2EC0$\nu$) the emission of two neutrinos~\cite{Primakoff_Rosen}.
Neutrinoless double electron capture is the hypothetical capture of two electrons without neutrino emission. It is a rare process that would unambiguously indicate new physics beyond the Standard Model if observed. The importance of 2EC0$\nu$ stems from its potential to demonstrate that the neutrino is its own antiparticle, a Majorana particle, and to provide a clear evidence of lepton number violation. Furthermore, for the light Majorana neutrino exchange mechanism, the half-life of 2EC0$\nu$ is inversely proportional to the square of the effective Majorana neutrino mass ($T_{1/2, 0\nu}^{-1}\sim (m^{\text{eff}}_{\nu})^2$)~\cite{Doi_Kotani}, offering a unique pathway to determination the absolute neutrino mass scale.
The theoretical study of 2EC0$\nu$ is linked with the study of its Standard Model-allowed analogue, 2EC2$\nu$. The half-life  for 2EC2$\nu$ is given by the following expression:
\begin{equation}
    (T_{1/2,\,2\nu})^{-1} \propto G^{2\nu}|M_{2\nu}|^2,
\end{equation}
where $G^{2\nu}$ is a known phase-space factor and $M_{2\nu}$ is a nuclear matrix element (NME)~\cite{SUHONEN1998123}. The NMEs for both the 2EC2$\nu$ and 2EC0$\nu$ processes share the same initial and final nuclear wave functions, but involve different transition operators and intermediate-state summations. Theoretical predictions for the 2EC2$\nu$ half-life are dominated by uncertainties in these NMEs. Experimentally measured half-lives thus provide a valuable test of nuclear models~\cite{NMEs, NMEs_Xe}. While agreement with 2EC2$\nu$ does not guarantee accurate 2EC0$\nu$ predictions, a point of ongoing debate, it serves as an important consistency check, lending credibility to neutrinoless mode calculations~\cite{Rodin2006}.

There are 34 candidate isotopes for which the 2EC2$\nu$ process has the shortest predicted half-lives. For 12 of these, including $^{36}$Ar considered in this paper, 2EC2$\nu$ mode is the only possible double electron capture channel (the $\varepsilon\beta^+, 2\beta^+, 2\beta^-$ mode are energetically forbidden)~\cite{Tretyak_tables}. Previously, the search for these processes was carried out on various isotopes, including $^{78}$Kr~\cite{Krypton, Krypton_2} and $^{124}$Xe~\cite{XENON1T, XENON1T_XENONnT}, as they are relatively accessible noble gas isotopes. 
The 2EC2$\nu$ process was first experimentally observed in $^{124}$Xe by the XENON1T experiment~\cite{XENON1T} and subsequently measured with increasing precision by XENONnT~\cite{XENONnT}, LZ~\cite{LZ}, and PandaX-4T~\cite{PandaX}.

The theoretical framework employed for calculating the NMEs in the 2EC2$\nu$ decay for $^{36}$Ar$\rightarrow^{36}$S is based on the realistic nuclear shell model with the USD (Universal \textit{sd}-shell Interaction) effective interaction~\cite{USD}. In this approach, the $^{36}$Ar nucleus is treated as an inert doubly magic $^{16}$O core (with filled proton and neutron shells), with valence nucleons in its outer \textit{sd}-shells: $0d_{5/2}, 1s_{1/2}$ and $0d_{3/2}$. It is the configurations of the valence nucleons in these orbitals that determine the properties of the low-lying states of the nucleus and the characteristics of nuclear transitions. Within this restricted model space, the Gamow–Teller transitions were computed including core-polarization and meson-exchange current corrections~\cite{Nakada1994}. Intermediate nuclear states were treated using Whitehead’s moment method -- a mathematical technique that ensures a stable result for the NME even when the strengths of individual transitions are not well-converged~\cite{Whitehead}. However, the model exhibits uncertainties due to: the restriction to the \textit{sd}-shell, neglecting higher-lying configurations; the omission of Coulomb interactions in the USD Hamiltonian; and the dependence on empirical Gamow-Teller operator corrections, which may introduce systematic uncertainties.

Therefore, the half-life of the 2EC2$\nu$ for $^{36}$Ar, obtained from theoretical calculations of the NMEs, is $T_{1/2}^{2\text{EC}2\nu} ({^{36}\text{Ar}}\rightarrow{^{36}\text{S}}) = \SI{1.7E29}{yr}$~\cite{theory}. 
The predicted half life for $^{36}$Ar is significantly higher than for $^{124}$Xe,  due to the smaller phase space factor and reduced NME. The phase space factor scales as $G^{2\nu}\sim Q^{5}$~\cite{SUHONEN1998123}, where $Q$ is decay energy. Since $Q(^{36}\text{Ar})\approx 433\text{ keV}$ is substantially  smaller than $Q(^{124}\text{Xe})\approx 2.8\text{ MeV}$, leading to $G^{2\nu}(^{36}\text{Ar})\ll G^{2\nu}(^{124}\text{Xe})$. Additionally, the double-magic nature of $^{36}$Ar ($Z=N=18$), leads to a suppressed NME compared to the non-magic $^{124}$Xe. In a non-magic nucleus ($Z=54,\: N=70$) the presence of unpaired nucleons creates favourable conditions for nuclear transitions. Moreover, the intermediate states of $^{124}$Te in this case turn out to be closer to resonance, significantly enhancing the contribution to the matrix element. These factors combine to make the NME for $^{124}$Xe about an order of magnitude larger than for $^{36}$Ar.

The successful  observation of 2EC2$\nu$ in $^{124}$Xe by XENON1T~\cite{XENON1T} experiment provides critical validation of theoretical models like Quasiparticle Random Phase Approximation (QRPA)~\cite{Vogel1986, SUHONEN1998123, Suhonen_Xe} and the shell models for 2EC2$\nu$ processes. This empirical confirmation suggests that such models can be reliably applied to $^{36}$Ar, despite the complexities involved in its nuclear matrix element calculations~\cite{theory}.

For $^{36}$Ar, dedicated searches have focused on the neutrinoless mode (2EC0$\nu$). The GERDA collaboration has established lower limits on its half-life of $3.6\times10^{21}$ years (Phase I)~\cite{GERDA2016} and $1.5\times10^{22}$ years (Phase II)~\cite{GERDA2024}.
As discussed above, the predicted half-life for the two-neutrino mode is well beyond the sensitivity of current detectors. Achieving sensitivity to such long half-lives would require large detector masses, high isotopic enrichment, and long exposure times.
The successful validation of the NME calculations by the $^{124}$Xe data implies that any positive detection of 2EC2$\nu$ in $^{36}$Ar by a next-generation experiment would be highly significant. Since the theoretically predicted rate is far below current and near-future sensitivities, any observed signal would suggest either a breakdown of the nuclear model for this specific isotope or  the presence of new physics beyond the Standard Model. This fundamental physical potential motivates the search for this process in next-generation experiments such as DarkSide-20k and ARGO.

The 2EC$2\nu$ process has not previously been experimentally studied in the argon isotope $^{36}$Ar. We report here on the first experimental search for double electron capture in $^{36}$Ar using the dataset from the DarkSide-50 experiment.

\section{Emission induced by the 2EC2$\nu$ reaction}
\label{sec:2}

\begin{figure}[h!]
    \centering
    \begin{tikzpicture}[
  particle/.style={thick},
  vertex/.style={circle, fill=black, inner sep=2pt},
  label distance=3mm,
  decoration={markings, mark=at position 0.5 with {\arrow{>}}},
  wavy/.style={decorate, decoration={snake, amplitude=2pt, segment length=5pt}}, 
  scale=1.2 
]
  \draw[particle, postaction={decorate}] (-3,1.2) -- (-1,1.2) node[very near start, above, sloped] {$p$};
  \draw[particle, postaction={decorate}] (-3,0.7) -- (-1,0.7) node[very near start, below] {$e^-_{\text{bound}}$};
  \node at (-2, 0) {(A, Z)};
  \draw[particle, postaction={decorate}] (-3,-0.7) -- (-1,-0.7) node[very near start, above, sloped] {$e^-_{\text{bound}}$};
  \draw[particle, postaction={decorate}] (-3,-1.2) -- (-1,-1.2) node[very near start, below] {$p$};

  \coordinate (center) at (-1,0);
  \def\angle{30}
  
   \draw[wavy] (-1.5,1.2) to[bend left=30] (-0.3,0.7) node[below, sloped] {$W^{-}$};
   \draw[wavy] (-1.5,-1.2) to[bend left=30] (-0.3,-0.7)node[below, sloped] {$W^{-}$};
    
  \draw[particle, postaction={decorate}] (-1,1.2) -- (1,1.2) node[very near end, above, sloped] {$n$};
  \draw[particle, postaction={decorate}] (-1,0.7) -- (1,0.7) node[very near end, below] {$\nu_{e}$};
  \node at (0, 0) {(A, Z-2)};
  \draw[particle, postaction={decorate}] (-1,-0.7) -- (1,-0.7) node[very near end, above, sloped] {$\nu_{e}$};
  \draw[particle, postaction={decorate}] (-1,-1.2) -- (1,-1.2) node[very near end, below] {$n$};

\end{tikzpicture}
    \caption{Feynman diagram for 2EC2$\nu$ process.}
    \label{fig:diagram}
\end{figure}
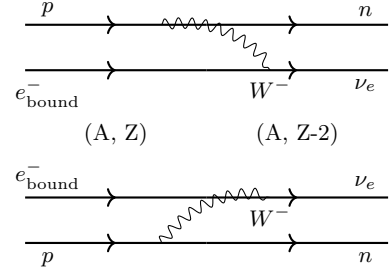

The energy released from the 2EC2$\nu$ is shared between the emitted neutrino pair and the rearrangement of vacancies in the atomic electron shells (Fig.~\ref{fig:diagram}). This makes the atom unstable and excited. The effective charge of the nucleus, perceived by the electrons in other shells, changes, which leads to a change in potential and, as a result, to the reconfiguration of the electron shell. This means that the electrons in other shells (\textit{L}, \textit{M}, etc.) end up in a slightly different energy state and distribution than in an ordinary atom~\cite{Krypton}. Electrons are predominantly captured from the \textit{K}-shell. This is not only because the \textit{K}-shell is spatially closest to the nucleus, but more fundamentally because the probability density of the \textit{K}-shell electrons at the nuclear site, $|\psi(0)|^2$, is the largest among those that are non-zero. The capture rate is directly proportional to this quantity, as the weak interaction is point-like. The same applies to higher shells, only the \textit{s}-subshells contribute, with probabilities decreasing rapidly from \textit{L} to \textit{M}, while capture from orbitals with $l>0$ (e.g. \textit{p}, \textit{d}) is essentially forbidden since their wavefunctions vanish at $r=0$.

In the process of double electron capture, the excitation of the atomic shell is taken off by a series of energy transitions of electrons from the upper shells to the lower ones. The released energy is carried away by one or more Auger electrons and/or one or more photons of characteristic radiation~\cite{electron_capture, electron_capture2, electron_capture3}. The challenge of observing double electron capture lies in detecting the associated atomic processes.
For $^{36}$Ar considered in this paper, the following processes take place:
\begin{equation}
    \begin{aligned}
        {^{36}\text{Ar}}+2e^- &\longrightarrow\: {^{36}\text{S}^{**}} + 2\nu_e \quad (Q \approx 433\ \text{keV})\\
        {^{36}\text{S}^{**}} &\longrightarrow {^{36}\text{S}}+n e^-_\text{Auger}+n' \gamma,
    \end{aligned}
\end{equation}
where $n,n'\geq 0$ are the numbers of emitted Auger electrons and characteristic photons.

The amount of energy release resulting from double electron capture in $^{36}$Ar was determined using the shake-off atomic model~\cite{Krypton}. It is important to note that the spectrum of these X-rays includes not only ordinary \textit{K}$_\alpha$ and \textit{K}$_\beta$ lines, but also hypersatellite and satellite lines shifted in energy due to changes in the binding energies of electrons in the inflated atomic shell. A hypersatellite is an X-ray photon emitted when an electron fills a vacancy in the \textit{K}-shell while two vacancies were initially present in that shell (i.e. the atom initially has a double \textit{K}‑hole), whereas satellite is an X-ray photon emitted when an electron fills a vacancy in the \textit{K}-shell, but the atom initially possesses at least one additional vacancy outside the K-shell (e.g. in the \textit{L} or \textit{M} shell) (Table~\ref{table:levels}). The extra vacancies in satellites and hypersatellites alter the screening of the nuclear charge, which shifts the emitted X‑ray energies relative to the ordinary lines.

\begin{figure}[h!]
    \centering
    \includegraphics[width=\linewidth]{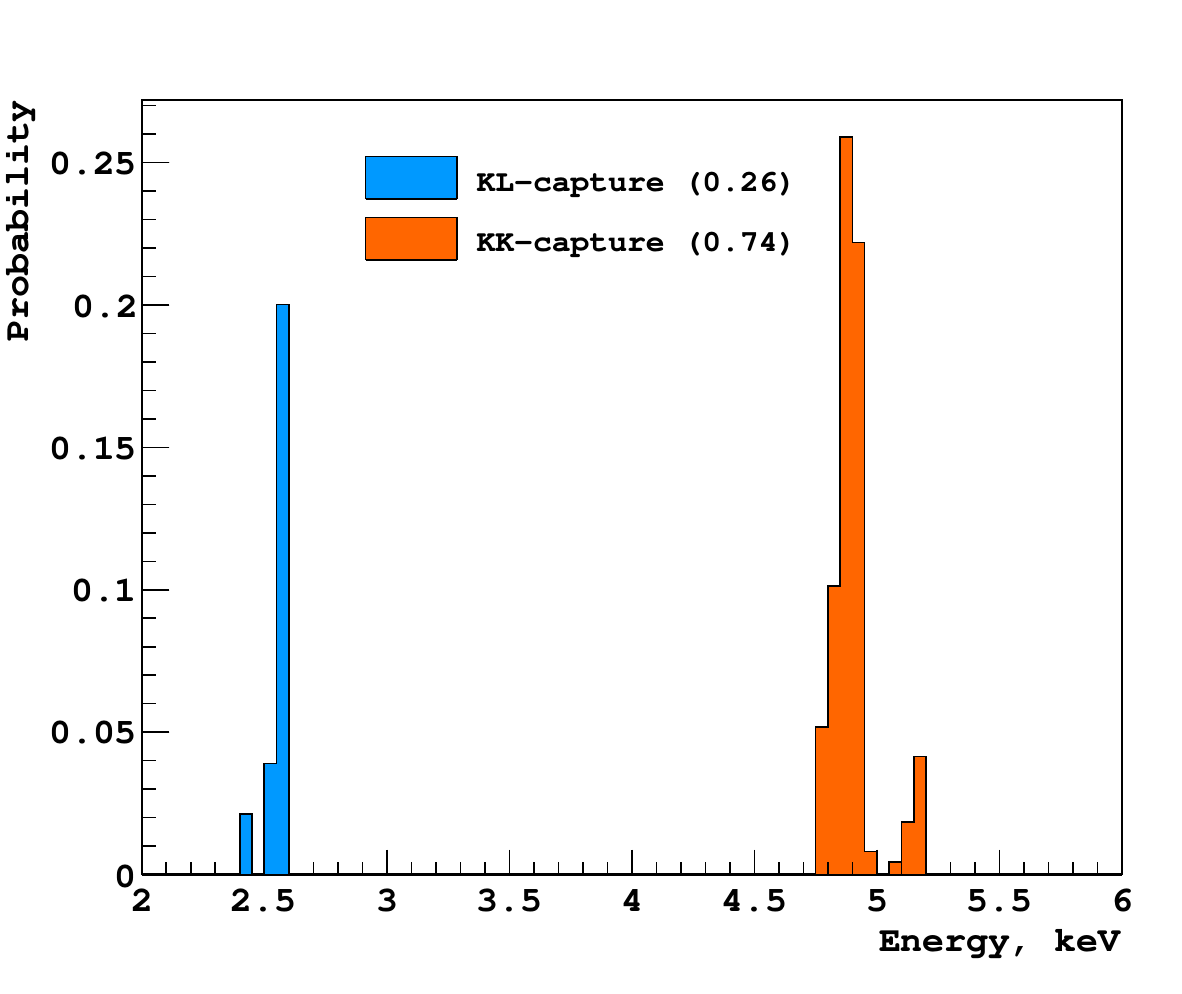}
    \captionof{figure}{Energy distribution of the 2EC2$\nu$ decay in $^{36}$Ar. The stacked histogram represents the absolute probabilities of the \textit{KK}-capture (orange) and \textit{KL}-capture (blue) channels. The \textit{KL} channel shows a peak at 2.59 keV (0.2), while the \textit{KK} channel distributes its probability across multiple peaks, with the highest at 4.91 keV (0.222).}
    \label{fig:prob_distr} 
\end{figure}

The probability of the \textit{KK}-holes and \textit{KL}-holes formation is $\sim$74\% and $\sim$26\%, respectively. The probability of double electron capture from the other shells is below 1\% and therefore negligible. Consequently, here we only considered the \textit{KK}-capture and \textit{KL}-capture. The related atomic shell energies and transition probabilities were calculated in the framework of the Dirac-Fock method, using the RAINE code package~\cite{RAINE_code}. The energy calculation accuracy is several eV. More details on the calculations can be found in Refs.~\cite{RAINE, Karpeshin}. 
The de-excitation of the excited atom proceeds in several successive stages: filling the first vacancy on the \textit{K}-shell (1st step), filling the second vacancy on the \textit{K}-shell/\textit{L}-shell (2nd step), filling the third vacancy on the \textit{L}-shell (3rd step) (see Tables~\ref{table:kk} and \ref{table:kl}, and Fig.~\ref{fig:prob_distr}). Auger electrons emitted during the atomic de-excitation cascade lose their energy primarily through ionization and excitation of surrounding argon atoms, producing a detectable ionization signal in the liquid argon medium. For the process initiated by the creation of a vacancy on the shell \textit{X}, with the transfer of an electron from the shell \textit{Y} to fill it and the emission of an Auger electron from the shell \textit{Z}, the kinetic energy of the electron is:
\begin{equation}
    E_{XYZ} = E_X(Z)-E_Y(Z)-E_Z(Z+1),
\end{equation}
where $E_X(Z)$ is the binding energy of an electron on the shell \textit{X} in the atom with atomic number \textit{Z}, $E_Y(Z)$ is the binding energy of an electron on the shell \textit{Y} and $E_Z(Z+1)$ is the binding energy of an electron on the shell \textit{Z} in an atom with atomic number $Z+1$.
Characteristic X-rays ($\gamma$-satellites) emitted during electron shell transitions undergo photoelectric absorption in argon atoms, accompanied by  photoelectron emission with energies equal to $E_\gamma - E_\text{binding}$, where $E_\text{binding}$ are summarised in Table~\ref{table:levels}.

\begin{table}[!h]
    \caption{The binding energies of the electrons in the $^{36}$Ar atom.}
    \label{table:levels}
    \begin{subtable}[h]{.5\linewidth}
      \caption{Hypersatellite transitions}
      \centering
        \begin{tabular}{lS[table-format=4.3]}
            \hline\noalign{\smallskip}
            {$NLJ$} & {$E$, eV} \\
            \noalign{\smallskip}\hline\noalign{\smallskip}
            1S+ & 2745.631 \\
            2S+ & 296.801 \\
            2P- & 245.649 \\
            2P+ & 244.929 \\
            3S+ & 32.173 \\
            3P- & 16.299 \\
            3P+ & 15.839 \\
        \end{tabular}
    \end{subtable}
    \begin{subtable}[h]{.45\linewidth}
      \centering
        \caption{Satellite transitions}
        \begin{tabular}{lS[table-format=4.3]}
            \hline\noalign{\smallskip}
            {$NLJ$} & {$E$, eV} \\
            \noalign{\smallskip}\hline\noalign{\smallskip}
            1S+ & 2583.284 \\
            2S+ & 283.324 \\
            2P- & 230.337 \\
            2P+ & 230.890 \\
            3S+ & 31.953 \\
            3P- & 16.119 \\
            3P+ & 15.370 \\
        \end{tabular}
    \end{subtable} 
\end{table}

\begin{table*}[!h]
    \centering
        \caption{The \textit{KK}-capture de-excitation channels at $^{36}$Ar}
    \label{table:kk}
    \begin{tabular}{cccccccc}
            \hline\noalign{\smallskip}
            $1$st step & $E_1$, eV & $P_1$ & 2nd step & $E_2$, eV & $P_2$ & 3rd step & $E_3$, eV\\
            \noalign{\smallskip}\hline\noalign{\smallskip}
             & & & $KL_{23}L_{23}$ & 2099 & 30\% & $n\times \text{LMM}$& $600$\\
            \noalign{\smallskip}\noalign{\smallskip}
            $KL_{23}L_{23}$\footnotemark & 2211 & 62\%  & $KL_{1}L_{23}$ & 2053 & 19\% & (150\text{ eV}) & $600$ \\
            \noalign{\smallskip}\noalign{\smallskip}
              &  &  & $KL_{1}L_{1}$ & 1996 & 6\% & & $600$\\ 
            \noalign{\smallskip}\noalign{\smallskip}
              &  &  & $\gamma\longrightarrow$ph.e & 2100 & 7\% & & $450$\\ 
             \noalign{\smallskip}\hline\noalign{\smallskip} 

             &  &  & $KL_{23}L_{23}$ & 2080 & 16\% & $n\times \text{LMM}$ & $600$\\
            \noalign{\smallskip}\noalign{\smallskip}
             $KL_{1}L_{23}$ & 2181 & 23.5\% & $KL_{1}L_{23}$ & 2024 & 3.9\% & (150\text{ eV}) & $600$\\
            \noalign{\smallskip}\noalign{\smallskip}
              &  &  & $\gamma\longrightarrow$ph.e & 2170 & 3.8\% & & $450$\\ 
             \noalign{\smallskip}\hline\noalign{\smallskip}  

             &  &  & $KL_{23}L_{23}$ & 2102 & 5.6\% & $n\times \text{LMM}$ & $600$\\
            \noalign{\smallskip}\noalign{\smallskip}
            $\gamma\longrightarrow$ph.e & 2470 & 9.4\%  & $KL_{1}L_{23}$ & 2048 & 2.5\% &(150\text{ eV}) &$600$\\
            \noalign{\smallskip}\noalign{\smallskip}
              &  &  & $KL_{1}L_{1}$ & 1985 & 0.6\% & &$600$\\ 
            \noalign{\smallskip}\noalign{\smallskip}
              &  &  & $\gamma\longrightarrow$ph.e & 2048 & 1.1\% & &$450$\\ 
             \noalign{\smallskip}\hline\noalign{\smallskip}  

            &  &  & $KL_{23}L_{23}$ & 2088 & 4.4\% & $n\times \text{LMM}$ &$450$\\
            \noalign{\smallskip}\noalign{\smallskip}
             $KL_{1}L_{1}$ & 2119 & 5.1\% & $\gamma\longrightarrow$ph.e & 2070 & 0.7\% & (150\text{ eV}) &$450$\\
        \end{tabular}
\end{table*}

\begin{table*}[!h]
    \centering
    \caption{The \textit{KL}-capture de-excitation channels at $^{36}$Ar}
    \label{table:kl}
    \begin{tabular}{lllll}
            \hline\noalign{\smallskip}
            $1$st step & $E_1$, eV & $P_1$ & 2nd step & $E_2$, eV \\
            \noalign{\smallskip}\hline\noalign{\smallskip}
            $KL_{23}L_{23}$ & 2107 & 77\% & $n\times$ \text{LMM} & 480 \\
            \noalign{\smallskip}\noalign{\smallskip}
            $KL_{1}L_{23}$ & 2045 & 15\% & $(160\text{ eV})$ & 480 \\
            \noalign{\smallskip}\noalign{\smallskip}
            $\gamma\longrightarrow$ph.e. & 2063 & 8.2\% & (178\text{ eV}) & 356 \\ 
            \noalign{\smallskip}\noalign{\smallskip}
        \end{tabular}
\end{table*}

\section{The DarkSide-50 experiment}
\label{sec:3}

The DarkSide-50 (DS-50) experiment was primarily built to search for spin-independent interactions between high-mass dark matter particles ($>\SI{10}{GeV/\it{c}^2}$) in the form of weakly interacting massive particles (WIMPs) and ato\-mic nuclei. Nevertheless, its analysis technique enabled the improvement of existing limits on WIMP-nucleon and WIMP-electron interactions in the few {GeV/$c^2$} and {sub-GeV/$c^2$} mass ranges in 2018 and subsequently in 2023 \cite{DarkSide:2018ppu, DarkSide:2018bpj, DarkSide:2022dhx, DarkSide:2022knj, DarkSide-50:2022qzh, DarkSide-50:2023fcw, 2025sensitivitylowmasswimpsimproved}.
The DarkSide-50 experiment was located at the Laboratori Na\-zio\-na\-li del Gran Sasso (LNGS) of the INFN, in Italy. The rock overburden of the underground laboratory is equal to \SI{3800}{\text{m.w.e.}} and it reduces the flux of cosmic muons by six orders of magnitude~\cite{underground}. 
The detector is a dual-phase liquid argon (LAr) time projection chamber (TPC) with an active mass of {$(46.4 \pm 0.7)$ kg}. Underground argon is used because it has lower concentration of $\beta$-radioactive $^{39}$Ar than atmospheric argon. Nevertheless, even this reduced amount requires a precise assessment to fully eliminate the background. Two arrays of 19 PMTs each (Hamamatsu R11065 3”)  detect photons, with one array positioned above the anode and the other below the cathode.

The DarkSide-50 TPC, surrounded by a double-walled stainless steel cryostat with vacuum insulation, is located inside a 30-ton liquid scintillator veto (LSV) for active neutron rejection. The LSV is equip\-ped with 110 8" PMTs, surrounded by a \SI{1}{kt} water Cherenkov veto with another 80 PMTs. The latter actively filters out cosmic muons and acts as a passive shield against external backgrounds. In this analysis, the active veto capabilities were not used, as studies showed that the resulting background reduction would be marginal for the signal region of interest.

Particle interactions in the active target induce scintillation pulses (S1 signal) and ionization electrons. The electrons drift upward in the applied electric field and reach the gas pocket, a small volume of gaseous argon located at the top of the TPC above the liquid argon. There, they produce a secondary light pulse via electroluminescence (S2 signal). Ultraviolet photons from S1 and S2 signals are converted into the visible range by a wavelength shifter -- tetraphenyl butadiene (TPB), which covers the inner surface of the TPC.
More details on the detector construction can be found in Refs.~\cite{TPC, DarkSide-50:2022qzh}.
\footnotetext{The standard Auger transition is denoted as \textit{XYZ}, where \textit{X} represents the initial vacancy shell, Y -- the filling electron shell, and \textit{Z} -- the emitted Auger electron shell. Subscripts (e.g., L$_{23}$) indicate permissible subshells.}

\section{Concentration of $^{36}$Ar in UAr}

The isotopic composition of UAr indicates its special origin. The most abundant isotope, $^{40}$Ar, is predominantly  radiogenic, formed during geological time as a result of the decay of $^{40}$K in the Earth's crust, with any primordial contribution being negligible. In contrast, the isotopes $^{36}$Ar and $^{38}$Ar are primordial; they were synthesized during stellar nucleosynthesis prior to the formation of the Solar System and were incorporated into the protosolar nebula~\cite{primordial40Ar, primordial_Mars}. As a result, these lighter isotopes are more abundant in the atmosphere, which accumulated gases from a larger reservoir of primordial material~\cite{Marty_2012}.
Also, a key difference between atmospheric and underground argon is the concentration of the radioactive isotope $^{39}$Ar (half-life of 302 years~\cite{DEAP_Ar39}). This isotope is continuously produced in the atmosphere by cosmic-ray-induced spallation reactions on stable $^{40}$Ar. In contrast, underground argon is shielded from cosmic rays by the rock overburden. This inherent radio-purity is a principal reason for using UAr in the DarkSide-50.
While the isotopic abundance of primordial $^{36}$Ar in atmospheric argon is 0.334\%~\cite{UAr_Allegre}, the focus of this work is on underground argon, where the $^{36}$Ar concentration is expected to be significantly depleted.

In order to determine the $^{36}$Ar abundance in UAr used for the DarkSide-50 target, we implemented an inductively coupled plasma mass spectrometry (ICP-MS) based procedure~\cite{ICP-MS}. This allows the isotopic composition of gases to be measured directly and was used to characterize a sample of the original UAr. This technique allows for the direct injection of argon gas into the mass spectrometer, without any chemical treatment, enabling rapid and sensitive detection of contaminants and isotopes by means of a fast and direct approach, complementary to other techniques~\cite{Santorelli:2023vzj}. We performed a relative measurement of the isotopic composition of argon by comparing the UAr sample to commercial atmospheric argon samples from Air Liquide under identical conditions. From this comparison, we determine the depletion factor $R_{^{36}\text{Ar}} = 45.6\pm0.8$, where the uncertainty is dominated by systematic effects arising from different configurations of the experimental setup (the statistical uncertainty for a given configuration is smaller). Using the known $^{36}$Ar abundance in atmospheric argon, $\eta_{^{36}\text{Ar}}^\text{AAr}=0.334\%$, the isotopic abundance of $^{36}$Ar in UAr is $\eta_{^{36}\text{Ar}}^\text{UAr} = \frac{\eta_{^{36}\text{Ar}}^\text{AAr}}{R_{^{36}\text{Ar}}} = (7.32\pm0.13)\times10^{-3}\%$. A more detailed description of the ICP-MS technique and isotopic analysis will be given in a dedicated paper.

\section{Data selection and background model}
\label{sec:data_selection}
This analysis was performed using 653.1 live-days of data, taken from December 12, 2015 to December 24, 2018 with UAr target. The total livetime used in the exposure calculation is 633.5 days, after accounting for a 3\% deadtime introduced by the spurious electron veto (a dedicated data-quality cut to suppress non-physical electron signals). The full energy deposit from 2EC2$\nu$ in $^{36}$Ar is about \SI{4.8}{keV}. The detector's response to particle interactions is measured in terms of the number of ionization electrons (N$_e$), which forms the fundamental observable energy scale. Through energy calibration, the keV units are converted into N$_e$ scale. For the relevant energy region, the calibration establishes the correspondence $[1, 10]~\text{keV}\approx[27, 100]~\text{N}_e$. Consequently, the region of interest for this 2EC2$\nu$ search overlaps with that of the main DarkSide-50 low-mass WIMP analysis (1–10 GeV/c$^2$), allowing the use of well-established data selection criteria, background models, and calibration procedures~\cite{DarkSide-50:2022qzh, response_calibration}.

The fiducial mass for this analysis is $(19.4 \pm 0.3)\text{ kg}$, about half of the total target mass. This volume excludes the outer region of the time projection chamber, which is more exposed to external radioactive backgrounds; only signals from the 7 central PMTs are used to define this inner fiducial volume. Events are counted only if the maximum fraction of the S2 charge is observed at one of the 7 central PMTs. The reconstruction of the XY coordinates is not used in this analysis, since the number of photoelectrons in this energy range is too small.

\section{Signal modeling and validation}

The energy scale for the low-mass analysis was calibrated using the $\beta$-decay spectrum of $^{39}$Ar and the \textit{L}-shell electron capture line of $^{37}$Ar. The \textit{L}-line's total energy release served as a benchmark for calibration in the low-energy region. The \textit{K}-shell capture in $^{37}$Ar was not used for calibration because its modeling is complicated by the potential overlap of ionization clouds from its subsequent atomic de-excitation cascade and recombination processes, introducing significant uncertainty. The same difficulty, accounting for overlapping ionization clouds from multiple electrons emitted in a short cascade, affects the modeling of the $^{36}$Ar 2EC2$\nu$ signal.

\begin{figure}[h!]
  \includegraphics[width=0.5\textwidth]{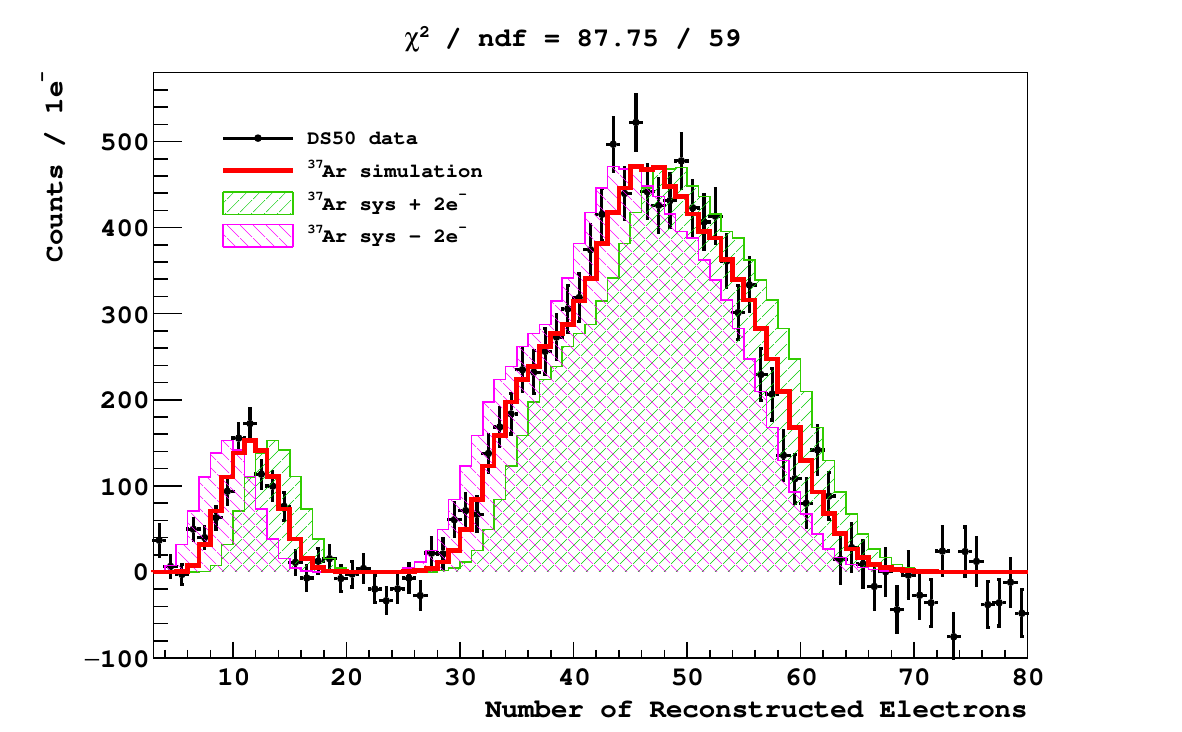}
  \caption{Detector response spectrum for the \textit{K}- and \textit{L}-captures in $^{37}$Ar. The black dots represent the experimental data obtained during the calibration of the detector after background subtraction. The red line indicates the detector response spectrum obtained from the total-energy approach described in the text. The light-green and magenta shaded histograms show the systematic variation of the simulated spectrum by $\pm2$ primary electrons (N$_e$), reflecting the uncertainty in modeling the cascade due to overlapping electron clouds.}
  \label{fig:ar37_spectrum}       
\end{figure}

To validate the response of the detector model to the 2EC2$\nu$ signal spectrum, we used the electron capture decays of $^{37}$Ar as a benchmark. The $^{37}$Ar nucleus captures an electron either from the \textit{K}-shell (90.4\% branching ratio, total energy releas 2.829 keV) or from the \textit{L}-shell (8.4\% branching ratio, 0.179 keV). The resulting atomic de-excitation cascades produce Auger electrons and X-rays, yielding an ionization signal in liquid argon similar to that expected from the $^{36}$Ar 2EC2$\nu$ decay.
The calibration data shown in Fig.~\ref{fig:ar37_spectrum} were obtained using a subtraction procedure described in Ref.~\cite{response_calibration}. Early data (first ~100 days), when $^{37}$Ar was present, and late data (subsequent 500 days), after $^{37}$Ar had almost entirely decayed, were each separately normalized by their respective livetimes. Subtracting the late-data spectrum from the early-data spectrum isolates the $^{37}$Ar signal by removing the time-independent background, primarily from residual $^{39}$Ar decays. This subtraction is the origin of the negative values visible on the \textit{y}-axis in Fig.~\ref{fig:ar37_spectrum}.

In Ref.~\cite{response_calibration}, this subtracted $^{37}$Ar spectrum was used to calibrate the ionization yield. For the \textit{L}-shell capture line, the contribution of primary Auger electrons (approximately 2.8 electrons per decay) was subtracted from the raw data during the calibration to extract the energy scale. For our validation, we have added back this contribution to recover the full signal expected from the $^{37}$Ar decay itself. For the \textit{K}-shell capture, the primary electrons were not included in the calibration because of the large overlap of electron clouds and the associated uncertainty from recombination in the cascade. The same limitation applies to our modeling; therefore, we do not attempt a detailed first-principles simulation of the full \textit{K}-shell cascade.

To validate our approach for the signal modeling, we tested two methods against the $^{37}$Ar \textit{K}-shell data. The first method attempted to simulate the full atomic cascade by treating each primary Auger electron and photoelectron as an independent energy deposit. This independent-electron approach failed to reproduce the measured $^{37}$Ar \textit{K}-shell peak, due to the fact, that the electrons are emitted in a very short time and the electron clouds caused by them overlap. The second method, treating the entire energy release of the cascade as a single, monoenergetic interaction point, reached a reasonable agreement with the data, as shown below. We therefore adopted this total-energy approach.

The \textit{K}-shell energy is 2.829 keV (90.4\%), and the \textit{L}-shell energy is 0.179 keV (8.4\%) accompanied by 2.8 primary electrons, giving $N_e (\text{\textit{K}-shell})\approx 48$ and $N_e (\text{\textit{L}-shell})\approx 11$. The black data points in Fig.~\ref{fig:ar37_spectrum} show the subtracted $^{37}$Ar spectrum from Ref.~\cite{response_calibration}, while the red line shows our simulated detector response using the total-energy method. The simulation reproduces both the position and the shape of the \textit{K}- and \textit{L}-shell peaks. A statistical evaluation of the agreement between the simulated and observed spectra yields $\chi^2/\text{ndf} = 87.75/59$ for the fit shown in Fig~\ref{fig:ar37_spectrum}, indicating that the residuals are well within the statistical uncertainties.
Nevertheless, minor discrepancies between the model and data remain. To conservatively account for the residual uncertainty in the energy scale, particularly relevant for the more complex multivacancy cascade of the $^{36}$Ar 2EC2$\nu$ process, we introduce a systematic shift of $\pm$2 primary electrons (N$_e$) in the simulated signal spectrum. As discussed below, varying the signal spectrum by this amount does not affect the final half-life limit.

Given this successful validation against $^{37}$Ar, we we employ a similar approach for the $^{36}$Ar 2EC2$\nu$ decay, calculating the detector response only for the total energy release of the \textit{KK}- and \textit{KL}-capture processes. A detailed, first-principles modeling of their individual complex de-excitation cascades is not feasible for the same reasons outlined for the $^{37}$Ar \textit{K}-shell capture.

The final dataset for this analysis contains approximately \num{3E5} events within a region of interest (ROI) of [4,170] N$_e$, corresponding to the energy range $[0.06, 21]$ keV$_{er}$. The 2EC2$\nu$ signal, as mentioned above, is contained within the [27, 100]~N$_e$ subrange, which is fully inside this ROI. The total exposure used in this analysis is $M\times T = 19.4\text{ kg}\times 633.5\text{ days}\approx12.3\text{ ton}\cdot\text{days}$. The main sources of background events in the energy ROI and fiducial volume are $^{39}$Ar and $^{85}$Kr decays occurring in the LAr bulk and $\gamma$s and X-rays from radioactive contaminants in the PMTs and stainless-steel cryostat. Background from radiogenic and cosmogenic neutrons as well as from the coherent elastic neutrino-nucleus scattering from solar and atmospheric neutrinos are negligible in our analysis. More details on the event selection and background models are provided in Ref.~\cite{DarkSide-50:2022qzh}.

The energy release and transition probabilities for \linebreak the 2EC2$\nu$ process in $^{36}$Ar were calculated using \linebreak the RAINE software package for atomic structure calculation~\cite{Karpeshin}, and reference material~\cite{RAINE}. As detailed in Sec.~\ref{sec:2}, these calculations account for all possible decay branches and their respective probabilities.

\begin{figure}[h!]
  \includegraphics[width=0.53\textwidth]{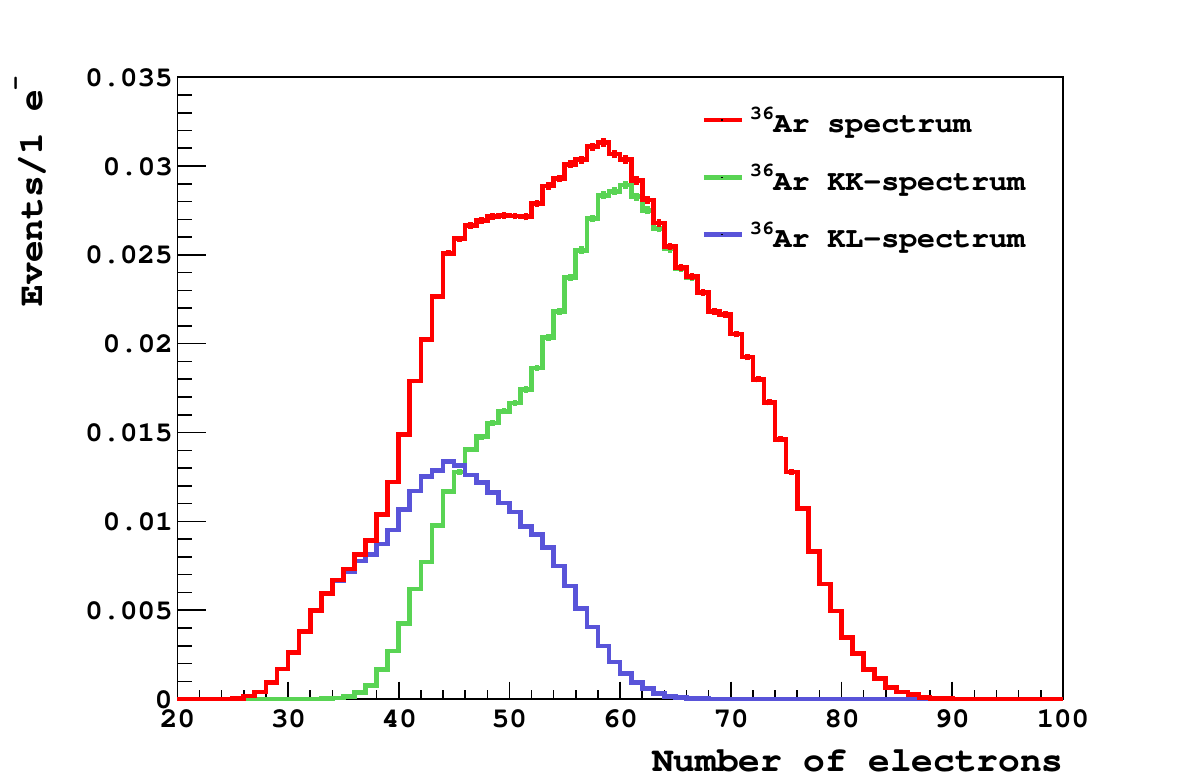}
  \caption{Simulated detector response spectrum for the 2EC2$\nu$ process, showing contributions from both \textit{KK}- and \textit{KL}-capture channels.}
  \label{fig:signal_spectrum}       
\end{figure}

The signal model for the $^{36}$Ar 2EC2$\nu$ was implemented as a composite spectrum of two monoenergetic lines. The energies of these lines, $\sim$4.9 keV for \textit{KK}-capture and $\sim$2.6 keV for \textit{KL}-capture were calculated as the probability-weighted sum of the total energy releases from all possible atomic de-excitation cascades following the respective captures (Fig.~\ref{fig:prob_distr}). These calculated energies were first converted into the expected number of ionization electrons (N$_e$), and the resulting spectrum was then convolved with the detector response function. The resulting simulated signal spectrum, derived from the analytical detector response function, exhibits distinct peaks at approximately 62 $N_e$ and 47 $N_e$ for \textit{KK}- and \textit{KL}-capture, respectively (Fig.~\ref{fig:signal_spectrum}).  The entire energy response of the 2EC2$\nu$ signal (99.9\%) falls within the range of [25, 90] N$_e$. Given the high statistics in this energy region, data are used in a binned form, with a 0.25 $N_e$ binning. 

\section{Data analysis}
\label{sec:6}

The analysis utilises a binned profile likelihood approach implemented through the RooFit/HistFactory statistical package~\cite{RooFit}. 
The methodology considers the content of each energy bin as an independent Poisson process, where the expected event count $\lambda$ in bin $i$ follows the relation $\lambda_i = \mu\cdot s_i + b_i$, with $\mu$ representing the signal strength parameter, $s_i$ the predicted weight of signal events from 2EC2$\nu$ process in bin $i$, and $b_i$  the estimated background contribution in the same energy bin.

In the unconditional maximum likelihood fit, $\mu$ is left unconstrained (i.e., the estimator $\hat{\mu}$ is allowed to take negative values), as required by the asymptotic formulae for the profile likelihood ratio~\cite{Cowan_2011}. The physical constraint $\mu\ge0$ is subsequently imposed when constructing the upper limit.
All nuisance parameters \(\theta\) relating to normalization and shape of background and signal contributions are constrained by Gaussian penalty terms. 
We use the same set of nuisance parameters as in Ref.~\cite{DarkSide-50:2022qzh}, complemented by the additional systematics related to signal modeling discussed in Sec.~\ref{sec:data_selection}, which have no impact on the final result.
The profile likelihood ratio test statistic for a given $\mu$ is defined as:
\begin{equation}
\label{eq:LLN}
    q_\mu = -2\ln[\mathcal{L}(\mu,\doublehat{\theta}_\mu)/\mathcal{L}(\hat{\mu},\hat{\theta})],
\end{equation}

where $\hat{\mu}$ is the unconditional maximum likelihood estimate of the signal strength, and $\doublehat{\theta}_{\mu}$ and $\hat{\theta}$ denote the conditional and unconditional maximum likelihood estimates of the nuisance parameters $\theta$, respectively.

For the purpose of upper limit calculation, the test statistic is modified following~\cite{Cowan_2011} to account for the physical boundary \(\mu \ge 0\):
\begin{equation}
\tilde{q}_{\mu} = \begin{cases}
q_{\mu}, & \hat{\mu} \le \mu, \\
0, & \hat{\mu} > \mu.
\end{cases}
\end{equation}
In this definition, the unconditional estimator $\hat{\mu}$ is allowed to take negative values during the fit, as required by the asymptotic approximation, while the reported upper limit is constrained to $\mu \ge 0$.

Initial validation of the background model was performed through background-only fits across the full $N_e$ range of [4, 170], which demonstrated satisfactory agreement with the observed spectrum~\cite{DarkSide-50:2022qzh}. 
The 90\% confidence level upper limit on $\mu$ was obtained by solving $(p_\mu = 0.10$ for $\mu$, where the $p$-value is given by the asymptotic formula $p_{\mu} = 1 - \Phi(\sqrt{\tilde{q}_{\mu}})$~\cite{Cowan_2011}.

\section{Results}
\label{sec:7}

The best fit value, determined by minimising the profiled likelihood ratio, yields $N_{\text{2EC2}\nu}=0$, indicating that no signal events were observed attributable to the 2EC2$\nu$ decay mode.
The upper limit on the number of such events, obtained at a 90\% confidence level (C.L.), is $N_{\text{2EC2}\nu} < 278$ events. The systematic uncertainty on the $^{36}$Ar abundance ($\pm1.8\%$ relative) as incorporated as a nuisance parameter in the likelihood fit. Its impact on the extracted upper limit was found to be negligible, and therefore does not affect the result.

\begin{figure}[h!]
  \includegraphics[width=0.5\textwidth]{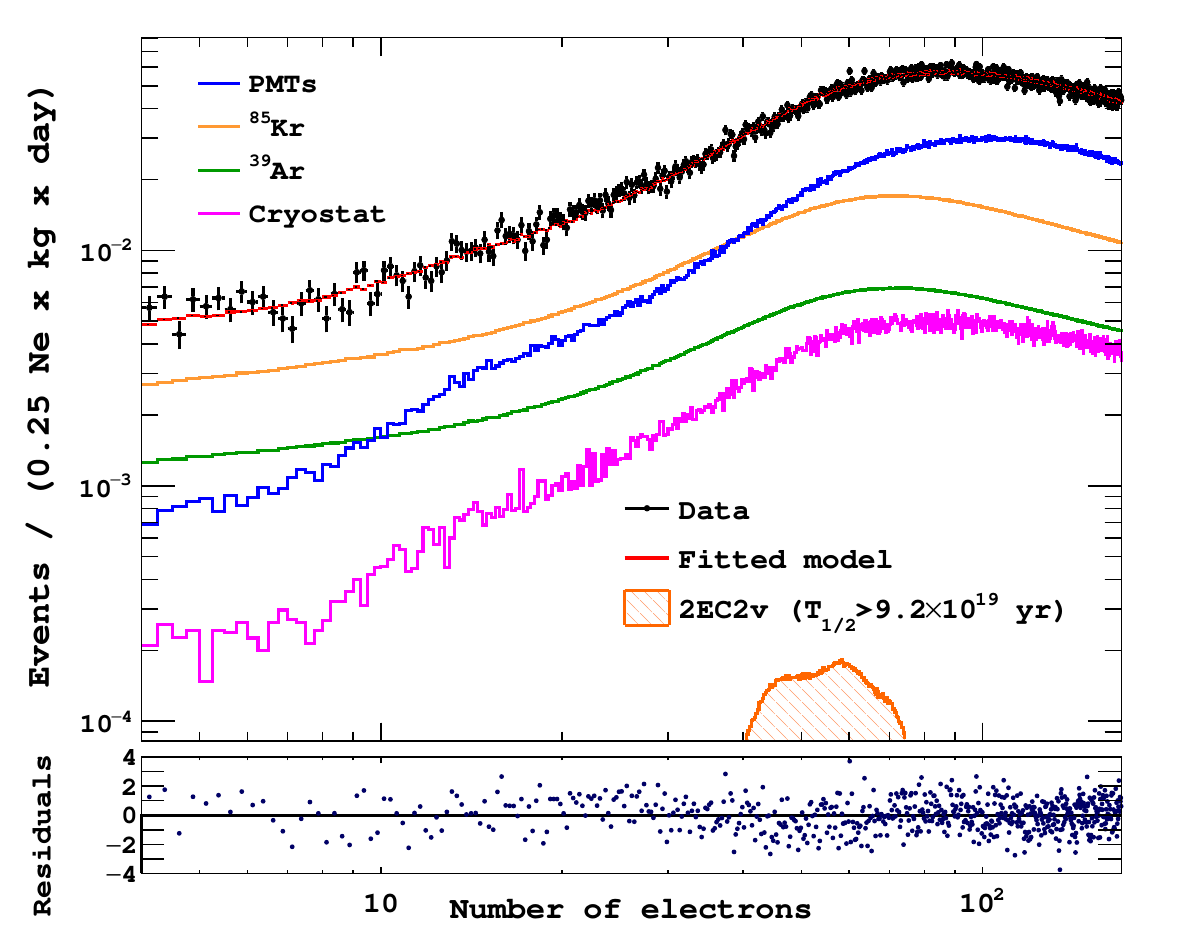}
  \caption{Data and background model compared to expected 2EC2$\nu$ spectra, and the individual contributions from the internal ($^{39}$Ar and $^{85}$Kr) and external components (cryostat and PMTs). The residuals are defined as the difference between the observed and expected events, normalized to the expected ones.}
  \label{fig:signal+bg_spectrum}       
\end{figure}

This upper limit can be translated into a lower limit on the half-life for the 2EC2$\nu$ process by
\begin{equation}
    T_{1/2}^{\text{2EC2}\nu}=\ln(2)\times\frac{N_A\times\eta^{\text{UAr}}_{^{36}\text{Ar}}}{N_{\text{2EC2}\nu}\times M_A}\times M \times T,
\end{equation}
where $M_A=\SI{0.04}{kg/mol}$ represents the molar mass of argon and $N_A$ denotes Avogadro’s constant. The isotopic abundance of the $^{36}$Ar isotope within the UAr is given by $\eta^{\text{UAr}}_{^{36}\text{Ar}} = \eta_{^{36}\text{Ar}}^{\text{AAr}}/R_{^{36}\text{Ar}}$, where $\eta_{^{36}\text{Ar}}^{\text{AAr}} = 0.334\%$ is the isotopic abundance of $^{36}$Ar in AAr, and $R_{^{36}\text{Ar}} = 45.6\pm0.8$ is the depletion factor of $^{36}$Ar abundance in UAr. The active mass of the liquid argon (LAr) target volume is givern by $M = \SI{19.4}{kg}$, and $T = 653.1\times0.97 =\SI{633.5}{days}$ represents the total livetime, after excluding the deadtime which accounts for 3\% of the raw exposure. Substituting the aforementioned parameters into the equation, the analysis of the DarkSide-50 experimental data provides a 90\% C.L. lower limit on the half-life of $T_{1/2}^{\text{2EC2}\nu}> 9.2 \times 10^{19}$ years for the $^{36}$Ar 2EC2$\nu$. Figure~\ref{fig:signal+bg_spectrum} shows the observed data together with the background model and the expected 2EC2$\nu$ signal spectrum corresponding to this half-life limit.

\section{Projected sensitivity with DarkSide-20k}

The sensitivity of the forthcoming DarkSide-20k~\cite{DS-20k} experiment will significantly improve upon the lower limit established in this work. The fiducial volume of the TPC within this setup will contain approximately \SI{34}{t} of UAr fiducial mass, representing an approximate three orders of magnitude increase compared to DarkSide-50. Furthermore, the DarkSide-20k experiment will employ Silicon Photomultipliers (SiPMs) instead of Photomultiplier Tubes (PMTs), offering a considerably reduced intrinsic radioactivity per unit mass and contributing to a $\sim$2.5-fold reduction in external $\gamma$-background per unit area. The TPC walls will be constructed from a polymethylmethacrylate (PMMA)-based material, exhibiting superior radio-purity characteristics compared to the stainless steel used in DarkSide-50. Crucially, both the DarkSide-50 and DarkSide-20k experiments utilise underground argon extracted from the same mine; consequently, the activity of $^{39}$Ar is assumed to be comparable in both detectors. Additionally, cryogenic distillation of the underground argon will be performed using the ARIA facility~\cite{ARIA}, thereby effectively suppressing residual traces of $^{85}$Kr by a factor of 100, which simplifies the background model to one dominated by a single $\beta$-source ($^{39}$Ar). These enhancements, together with improved data collection efficiency, an enlarged fiducial volume, and the potential for an analysis with a lower energy threshold enabled by better control of spurious electrons~\cite{spurious_electrons}, render the DarkSide-20k experiment significantly more sensitive to rare decay processes such as two-neutrino double electron capture~\cite{DarkSide-50:2022qzh, DS20k_sensitivity}.

The increased exposure results in a corresponding increase in the number of registered events. According to the law of large numbers, the variance of background fluctuations scales linearly with the number of events, i.e. $\sigma \propto \sqrt{N}$. 
Furthermore, the reduction of the specific background rate by a factor of $\sim$20 (due to $^{85}$Kr suppression and radiopure materials) provides an additional gain in sensitivity of $\sqrt{20} \approx 4.5$. 
Consequently, in the absence of systematic uncertainties, the experimental sensitivity scales as $\propto \sqrt{\mathcal{E}}$, where $\mathcal{E}$ denotes the total exposure. 
DarkSide-20k foresees an improvement of approximately one order of magnitude in sensitivity to WIMP interactions compared to DarkSide-50 in only one year of exposure. 
By the same scaling argument, combining the factors of increased mass ($\sim$1000x), reduced background ($\sim$20x), and longer runtime, the sensitivity of DarkSide-20k to $2\text{EC}2\nu$ is expected to increase by two to three orders of magnitude at the same $^{36}$Ar concentration in a 10-year run.

\section{Conclusion}
\label{sec:8}

We have conducted a search for the 2EC2$\nu$ process in $^{36}$Ar employing the full exposure of the DarkSide-50 experiment. With no statistically significant excess of events observed above the anticipated background, we have established a new lower limit on the half-life of this rare decay, $T_{1/2}>9.2 \times 10^{19}$ years (90\% C.L.). This constitutes the first experimental constraint on the 2EC2$\nu$ in $^{36}$Ar. Furthermore, we have assessed the projected sensitivity of the forthcoming DarkSide-20k experiment, anticipating that it will be capable of probing half-lives up to about two orders of magnitude longer, i.e. $\sim10^{22}$ years, assuming a total exposure of 200 t$\cdot$yr. Beyond DarkSide-20k, future multi-tonne scale experiments, such as those envisioned within the Global Argon Dark Matter Collaboration (GADMC)(e.g., ARGO), hold the potential to extend sensitivity to even longer half-lives.

\section{Acknowledgements}

The DarkSide Collaboration offers its deep gratitude to LNGS and its staff for their invaluable technical and logistical support. 
The authors also thank the Fermilab Particle Physics, Scientific, and Core Computing Divisions.
 Construction and operation of the DarkSide-50 detector was supported by the U.S. National Science Foundation (NSF) (Grants No. PHY-0919363, No. PHY-1004072, No. PHY-1004054, No. PHY-1242585, No. PHY-1314483, No. PHY-1314501, No. PHY-1314507, No. PHY-1352795, No. PHY-1622415, and associated collaborative Grants No. PHY-1211308 and No. PHY-1455351), 
 the Italian Istituto Nazionale di Fisica Nucleare, 
 the U.S. Department of Energy (Contracts No. DE-FG02-91ER40671, No. DEAC02-07CH11359, and No. DE-AC05-76RL01830), the Polish NCN (Grant No. UMO-2023/51/B/ST2/02099) 
 and the Polish Ministry for Education and Science (Grant No. 6811/IA/SP/2018). 
 We also acknowledge financial support from the French Institut National de Physique Nucléaire et de Physique des Particules (IN2P3), the IN2P3-COPIN consortium (Grant No. 20-152), and the UnivEarthS LabEx program (Grants No. ANR-10-LABX-0023 and No. ANR-18-IDEX-0001), 
 from the S{\~a}o Paulo Research Foundation (FAPESP) (Grants  No. 2021/11489-7), 
 from the Interdisciplinary Scientific and Educational School of Moscow University “Fundamental and Applied Space Research,” 
 from the Program of the Ministry of Education and Science of the Russian Federation for higher education establishments, Project No. FZWG-2020-0032 (2019-1569), 
 the International Research Agenda Programme AstroCeNT (MAB/2018/7) funded by the Foundation for Polish Science from the European Regional Development Fund, 
 and the European Union’s Horizon 2020 research and innovation program under Grant Agreement No. 952480 (DarkWave), the National Science Centre, Poland (2021/42/E/ST2/00331), 
 and from the Science and Technology Facilities Council, United Kingdom.
  I. Albuquerque was partially supported by the Brazilian Research Council (CNPq).
  The research has been supported by Theoretical Physics and Mathematics Advancement Foundation "BASIS" under Grant No. 23-22-5-1.
   This work was supported by the Spanish Ministry of Science and Innovation (MICINN) under Grant PID2022-138357NB-C22.


\begin{thebibliography}{52}%
\makeatletter
\providecommand \@ifxundefined [1]{%
 \@ifx{#1\undefined}
}%
\providecommand \@ifnum [1]{%
 \ifnum #1\expandafter \@firstoftwo
 \else \expandafter \@secondoftwo
 \fi
}%
\providecommand \@ifx [1]{%
 \ifx #1\expandafter \@firstoftwo
 \else \expandafter \@secondoftwo
 \fi
}%
\providecommand \natexlab [1]{#1}%
\providecommand \enquote  [1]{``#1''}%
\providecommand \bibnamefont  [1]{#1}%
\providecommand \bibfnamefont [1]{#1}%
\providecommand \citenamefont [1]{#1}%
\providecommand \href@noop [0]{\@secondoftwo}%
\providecommand \href [0]{\begingroup \@sanitize@url \@href}%
\providecommand \@href[1]{\@@startlink{#1}\@@href}%
\providecommand \@@href[1]{\endgroup#1\@@endlink}%
\providecommand \@sanitize@url [0]{\catcode `\\12\catcode `\$12\catcode `\&12\catcode `\#12\catcode `\^12\catcode `\_12\catcode `\%12\relax}%
\providecommand \@@startlink[1]{}%
\providecommand \@@endlink[0]{}%
\providecommand \url  [0]{\begingroup\@sanitize@url \@url }%
\providecommand \@url [1]{\endgroup\@href {#1}{\urlprefix }}%
\providecommand \urlprefix  [0]{URL }%
\providecommand \Eprint [0]{\href }%
\providecommand \doibase [0]{https://doi.org/}%
\providecommand \selectlanguage [0]{\@gobble}%
\providecommand \bibinfo  [0]{\@secondoftwo}%
\providecommand \bibfield  [0]{\@secondoftwo}%
\providecommand \translation [1]{[#1]}%
\providecommand \BibitemOpen [0]{}%
\providecommand \bibitemStop [0]{}%
\providecommand \bibitemNoStop [0]{.\EOS\space}%
\providecommand \EOS [0]{\spacefactor3000\relax}%
\providecommand \BibitemShut  [1]{\csname bibitem#1\endcsname}%
\let\auto@bib@innerbib\@empty
\bibitem [{\citenamefont {Winter}(1955)}]{Winter_theory}%
  \BibitemOpen
  \bibfield  {author} {\bibinfo {author} {\bibfnamefont {R.~G.}\ \bibnamefont {Winter}},\ }\href {https://doi.org/10.1103/PhysRev.100.142} {\bibfield  {journal} {\bibinfo  {journal} {Phys. Rev.}\ }\textbf {\bibinfo {volume} {100}},\ \bibinfo {pages} {142} (\bibinfo {year} {1955})}\BibitemShut {NoStop}%
\bibitem [{\citenamefont {Primakoff}\ and\ \citenamefont {Rosen}(1959)}]{Primakoff_Rosen}%
  \BibitemOpen
  \bibfield  {author} {\bibinfo {author} {\bibfnamefont {H.}~\bibnamefont {Primakoff}}\ and\ \bibinfo {author} {\bibfnamefont {S.~P.}\ \bibnamefont {Rosen}},\ }\href {https://doi.org/10.1088/0034-4885/22/1/305} {\bibfield  {journal} {\bibinfo  {journal} {Rep. Prog. Phys.}\ }\textbf {\bibinfo {volume} {22}},\ \bibinfo {pages} {121} (\bibinfo {year} {1959})}\BibitemShut {NoStop}%
\bibitem [{\citenamefont {Doi}\ \emph {et~al.}(1985)\citenamefont {Doi}, \citenamefont {Kotani},\ and\ \citenamefont {Takasugi}}]{Doi_Kotani}%
  \BibitemOpen
  \bibfield  {author} {\bibinfo {author} {\bibfnamefont {M.}~\bibnamefont {Doi}}, \bibinfo {author} {\bibfnamefont {T.}~\bibnamefont {Kotani}},\ and\ \bibinfo {author} {\bibfnamefont {E.}~\bibnamefont {Takasugi}},\ }\href {https://doi.org/10.1143/PTPS.83.1} {\bibfield  {journal} {\bibinfo  {journal} {Prog. Theor. Phys. Suppl.}\ }\textbf {\bibinfo {volume} {83}},\ \bibinfo {pages} {1} (\bibinfo {year} {1985})}\BibitemShut {NoStop}%
\bibitem [{\citenamefont {Suhonen}\ and\ \citenamefont {Civitarese}(1998)}]{SUHONEN1998123}%
  \BibitemOpen
  \bibfield  {author} {\bibinfo {author} {\bibfnamefont {J.}~\bibnamefont {Suhonen}}\ and\ \bibinfo {author} {\bibfnamefont {O.}~\bibnamefont {Civitarese}},\ }\href {https://doi.org/10.1016/S0370-1573(97)00087-2} {\bibfield  {journal} {\bibinfo  {journal} {Phys. Rep.}\ }\textbf {\bibinfo {volume} {300}},\ \bibinfo {pages} {123} (\bibinfo {year} {1998})}\BibitemShut {NoStop}%
\bibitem [{\citenamefont {Ni\c{t}escu}\ \emph {et~al.}(2024{\natexlab{a}})\citenamefont {Ni\c{t}escu}, \citenamefont {Ghinescu}, \citenamefont {Sevestrean} \emph {et~al.}}]{NMEs}%
  \BibitemOpen
  \bibfield  {author} {\bibinfo {author} {\bibfnamefont {O.}~\bibnamefont {Ni\c{t}escu}}, \bibinfo {author} {\bibfnamefont {S.}~\bibnamefont {Ghinescu}}, \bibinfo {author} {\bibfnamefont {V.~A.}\ \bibnamefont {Sevestrean}}, \emph {et~al.},\ }\href {https://doi.org/10.3390/universe10020098} {\bibfield  {journal} {\bibinfo  {journal} {Universe}\ }\textbf {\bibinfo {volume} {10}},\ \bibinfo {pages} {98} (\bibinfo {year} {2024}{\natexlab{a}})}\BibitemShut {NoStop}%
\bibitem [{\citenamefont {Ni\c{t}escu}\ \emph {et~al.}(2024{\natexlab{b}})\citenamefont {Ni\c{t}escu}, \citenamefont {Ghinescu}, \citenamefont {Sevestrean} \emph {et~al.}}]{NMEs_Xe}%
  \BibitemOpen
  \bibfield  {author} {\bibinfo {author} {\bibfnamefont {O.}~\bibnamefont {Ni\c{t}escu}}, \bibinfo {author} {\bibfnamefont {S.}~\bibnamefont {Ghinescu}}, \bibinfo {author} {\bibfnamefont {V.~A.}\ \bibnamefont {Sevestrean}}, \emph {et~al.},\ }\href {https://doi.org/10.1088/1361-6471/ad8767} {\bibfield  {journal} {\bibinfo  {journal} {J. Phys. G}\ }\textbf {\bibinfo {volume} {51}},\ \bibinfo {pages} {125103} (\bibinfo {year} {2024}{\natexlab{b}})}\BibitemShut {NoStop}%
\bibitem [{\citenamefont {Rodin}\ \emph {et~al.}(2006)\citenamefont {Rodin}, \citenamefont {Faessler}, \citenamefont {{\v{S}}imkovic},\ and\ \citenamefont {Vogel}}]{Rodin2006}%
  \BibitemOpen
  \bibfield  {author} {\bibinfo {author} {\bibfnamefont {V.}~\bibnamefont {Rodin}}, \bibinfo {author} {\bibfnamefont {A.}~\bibnamefont {Faessler}}, \bibinfo {author} {\bibfnamefont {F.}~\bibnamefont {{\v{S}}imkovic}},\ and\ \bibinfo {author} {\bibfnamefont {P.}~\bibnamefont {Vogel}},\ }\href {https://doi.org/10.1016/j.nuclphysa.2005.12.004} {\bibfield  {journal} {\bibinfo  {journal} {Nuclear Physics A}\ }\textbf {\bibinfo {volume} {766}},\ \bibinfo {pages} {107} (\bibinfo {year} {2006})}\BibitemShut {NoStop}%
\bibitem [{\citenamefont {Tretyak}\ and\ \citenamefont {Zdesenko}(2002)}]{Tretyak_tables}%
  \BibitemOpen
  \bibfield  {author} {\bibinfo {author} {\bibfnamefont {V.~I.}\ \bibnamefont {Tretyak}}\ and\ \bibinfo {author} {\bibfnamefont {Y.~G.}\ \bibnamefont {Zdesenko}},\ }\href {https://doi.org/10.1006/adnd.2001.0873} {\bibfield  {journal} {\bibinfo  {journal} {At. Data Nucl. Data Tables}\ }\textbf {\bibinfo {volume} {80}},\ \bibinfo {pages} {83} (\bibinfo {year} {2002})}\BibitemShut {NoStop}%
\bibitem [{\citenamefont {Ratkevich}\ \emph {et~al.}(2017)\citenamefont {Ratkevich}, \citenamefont {Gangapshev}, \citenamefont {Gavrilyuk} \emph {et~al.}}]{Krypton}%
  \BibitemOpen
  \bibfield  {author} {\bibinfo {author} {\bibfnamefont {S.~S.}\ \bibnamefont {Ratkevich}}, \bibinfo {author} {\bibfnamefont {A.~M.}\ \bibnamefont {Gangapshev}}, \bibinfo {author} {\bibfnamefont {Y.~M.}\ \bibnamefont {Gavrilyuk}}, \emph {et~al.},\ }\href {https://doi.org/10.1103/PhysRevC.96.065502} {\bibfield  {journal} {\bibinfo  {journal} {Phys. Rev. C}\ }\textbf {\bibinfo {volume} {96}},\ \bibinfo {pages} {065502} (\bibinfo {year} {2017})}\BibitemShut {NoStop}%
\bibitem [{\citenamefont {Gavrilyuk}\ \emph {et~al.}(2013)\citenamefont {Gavrilyuk}, \citenamefont {Gangapshev}, \citenamefont {Kazalov} \emph {et~al.}}]{Krypton_2}%
  \BibitemOpen
  \bibfield  {author} {\bibinfo {author} {\bibfnamefont {Y.~M.}\ \bibnamefont {Gavrilyuk}}, \bibinfo {author} {\bibfnamefont {A.~M.}\ \bibnamefont {Gangapshev}}, \bibinfo {author} {\bibfnamefont {V.~V.}\ \bibnamefont {Kazalov}}, \emph {et~al.},\ }\href {https://doi.org/10.1103/PhysRevC.87.035501} {\bibfield  {journal} {\bibinfo  {journal} {Phys. Rev. C}\ }\textbf {\bibinfo {volume} {87}},\ \bibinfo {pages} {035501} (\bibinfo {year} {2013})}\BibitemShut {NoStop}%
\bibitem [{\citenamefont {Aprile}\ \emph {et~al.}(2019)\citenamefont {Aprile} \emph {et~al.}}]{XENON1T}%
  \BibitemOpen
  \bibfield  {author} {\bibinfo {author} {\bibfnamefont {E.}~\bibnamefont {Aprile}} \emph {et~al.} (\bibinfo {collaboration} {XENON Collaboration}),\ }\href {https://doi.org/10.1038/s41586-019-1124-4} {\bibfield  {journal} {\bibinfo  {journal} {Nature}\ }\textbf {\bibinfo {volume} {568}},\ \bibinfo {pages} {532} (\bibinfo {year} {2019})},\ \bibinfo {note} {arXiv:1904.11002 [nucl-ex]}\BibitemShut {NoStop}%
\bibitem [{\citenamefont {Aprile}\ \emph {et~al.}(2022{\natexlab{a}})\citenamefont {Aprile} \emph {et~al.}}]{XENON1T_XENONnT}%
  \BibitemOpen
  \bibfield  {author} {\bibinfo {author} {\bibfnamefont {E.}~\bibnamefont {Aprile}} \emph {et~al.} (\bibinfo {collaboration} {XENON Collaboration}),\ }\href {https://doi.org/10.1103/PhysRevC.106.024328} {\bibfield  {journal} {\bibinfo  {journal} {Phys. Rev. C}\ }\textbf {\bibinfo {volume} {106}},\ \bibinfo {pages} {024328} (\bibinfo {year} {2022}{\natexlab{a}})},\ \bibinfo {note} {arXiv:2205.04158 [hep-ex]}\BibitemShut {NoStop}%
\bibitem [{\citenamefont {Aprile}\ \emph {et~al.}(2022{\natexlab{b}})\citenamefont {Aprile} \emph {et~al.}}]{XENONnT}%
  \BibitemOpen
  \bibfield  {author} {\bibinfo {author} {\bibfnamefont {E.}~\bibnamefont {Aprile}} \emph {et~al.} (\bibinfo {collaboration} {XENON Collaboration}),\ }\href {https://doi.org/10.1103/PhysRevLett.129.161805} {\bibfield  {journal} {\bibinfo  {journal} {Phys. Rev. Lett.}\ }\textbf {\bibinfo {volume} {129}},\ \bibinfo {pages} {161805} (\bibinfo {year} {2022}{\natexlab{b}})}\BibitemShut {NoStop}%
\bibitem [{\citenamefont {Aalbers}\ \emph {et~al.}(2024)\citenamefont {Aalbers} \emph {et~al.}}]{LZ}%
  \BibitemOpen
  \bibfield  {author} {\bibinfo {author} {\bibfnamefont {J.}~\bibnamefont {Aalbers}} \emph {et~al.} (\bibinfo {collaboration} {The LZ Collaboration}),\ }\href {https://doi.org/10.1088/1361-6471/ad9039} {\bibfield  {journal} {\bibinfo  {journal} {J. Phys. G}\ }\textbf {\bibinfo {volume} {52}},\ \bibinfo {pages} {015103} (\bibinfo {year} {2024})}\BibitemShut {NoStop}%
\bibitem [{\citenamefont {Bo}\ \emph {et~al.}(2025)\citenamefont {Bo} \emph {et~al.}}]{PandaX}%
  \BibitemOpen
  \bibfield  {author} {\bibinfo {author} {\bibfnamefont {Z.}~\bibnamefont {Bo}} \emph {et~al.} (\bibinfo {collaboration} {PandaX-4T, PandaX Collaborations}),\ }\href {https://doi.org/10.1007/JHEP05(2025)119} {\bibfield  {journal} {\bibinfo  {journal} {JHEP}\ }\textbf {\bibinfo {volume} {2025}}\bibinfo  {number} { (5)},\ \bibinfo {pages} {119}}\BibitemShut {NoStop}%
\bibitem [{\citenamefont {Brown}\ and\ \citenamefont {Wildenthal}(2003)}]{USD}%
  \BibitemOpen
\bibfield  {number} {  }\bibfield  {author} {\bibinfo {author} {\bibfnamefont {B.}~\bibnamefont {Brown}}\ and\ \bibinfo {author} {\bibfnamefont {B.}~\bibnamefont {Wildenthal}},\ }\href {https://doi.org/10.1146/annurev.ns.38.120188.000333} {\bibfield  {journal} {\bibinfo  {journal} {Annu. Rev. Nucl. Part. Sci.}\ }\textbf {\bibinfo {volume} {38}},\ \bibinfo {pages} {29} (\bibinfo {year} {2003})}\BibitemShut {NoStop}%
\bibitem [{\citenamefont {Nakada}\ \emph {et~al.}(1994)\citenamefont {Nakada}, \citenamefont {Sebe},\ and\ \citenamefont {Otsuka}}]{Nakada1994}%
  \BibitemOpen
  \bibfield  {author} {\bibinfo {author} {\bibfnamefont {H.}~\bibnamefont {Nakada}}, \bibinfo {author} {\bibfnamefont {T.}~\bibnamefont {Sebe}},\ and\ \bibinfo {author} {\bibfnamefont {T.}~\bibnamefont {Otsuka}},\ }\href {https://doi.org/10.1016/0375-9474(94)90222-4} {\bibfield  {journal} {\bibinfo  {journal} {Nucl. Phys. A}\ }\textbf {\bibinfo {volume} {571}},\ \bibinfo {pages} {467} (\bibinfo {year} {1994})}\BibitemShut {NoStop}%
\bibitem [{\citenamefont {Whitehead}(1980)}]{Whitehead}%
  \BibitemOpen
  \bibfield  {author} {\bibinfo {author} {\bibfnamefont {R.~R.}\ \bibnamefont {Whitehead}},\ }\bibinfo {title} {Moment {M}ethods and {L}anczos {M}ethods},\ in\ \href {https://doi.org/10.1007/978-1-4613-3120-9_13} {\emph {\bibinfo {booktitle} {Theory and Applications of Moment Methods in Many-Fermion Systems}}},\ \bibinfo {editor} {edited by\ \bibinfo {editor} {\bibfnamefont {B.~J.}\ \bibnamefont {Dalton}}, \bibinfo {editor} {\bibfnamefont {S.~M.}\ \bibnamefont {Grimes}}, \bibinfo {editor} {\bibfnamefont {J.~P.}\ \bibnamefont {Vary}},\ and\ \bibinfo {editor} {\bibfnamefont {S.~A.}\ \bibnamefont {Williams}}}\ (\bibinfo  {publisher} {Springer},\ \bibinfo {address} {Boston},\ \bibinfo {year} {1980})\ pp.\ \bibinfo {pages} {235--255}\BibitemShut {NoStop}%
\bibitem [{\citenamefont {Nakada}\ \emph {et~al.}(1996)\citenamefont {Nakada}, \citenamefont {Sebe},\ and\ \citenamefont {Muto}}]{theory}%
  \BibitemOpen
  \bibfield  {author} {\bibinfo {author} {\bibfnamefont {H.}~\bibnamefont {Nakada}}, \bibinfo {author} {\bibfnamefont {T.}~\bibnamefont {Sebe}},\ and\ \bibinfo {author} {\bibfnamefont {K.}~\bibnamefont {Muto}},\ }\href {https://doi.org/10.1016/0375-9474(96)00227-8} {\bibfield  {journal} {\bibinfo  {journal} {Nucl. Phys. A}\ }\textbf {\bibinfo {volume} {607}},\ \bibinfo {pages} {235} (\bibinfo {year} {1996})}\BibitemShut {NoStop}%
\bibitem [{\citenamefont {Vogel}\ and\ \citenamefont {Zirnbauer}(1986)}]{Vogel1986}%
  \BibitemOpen
  \bibfield  {author} {\bibinfo {author} {\bibfnamefont {P.}~\bibnamefont {Vogel}}\ and\ \bibinfo {author} {\bibfnamefont {M.~R.}\ \bibnamefont {Zirnbauer}},\ }\href {https://doi.org/10.1103/PhysRevLett.57.3148} {\bibfield  {journal} {\bibinfo  {journal} {Phys. Rev. Lett.}\ }\textbf {\bibinfo {volume} {57}},\ \bibinfo {pages} {3148} (\bibinfo {year} {1986})}\BibitemShut {NoStop}%
\bibitem [{\citenamefont {Suhonen}(2013)}]{Suhonen_Xe}%
  \BibitemOpen
  \bibfield  {author} {\bibinfo {author} {\bibfnamefont {J.}~\bibnamefont {Suhonen}},\ }\href {https://doi.org/10.1088/0954-3899/40/7/075102} {\bibfield  {journal} {\bibinfo  {journal} {J. Phys. G}\ }\textbf {\bibinfo {volume} {40}},\ \bibinfo {pages} {075102} (\bibinfo {year} {2013})}\BibitemShut {NoStop}%
\bibitem [{\citenamefont {Agostini}\ \emph {et~al.}(2016)\citenamefont {Agostini} \emph {et~al.}}]{GERDA2016}%
  \BibitemOpen
  \bibfield  {author} {\bibinfo {author} {\bibfnamefont {M.}~\bibnamefont {Agostini}} \emph {et~al.} (\bibinfo {collaboration} {GERDA Collaboration}),\ }\href {https://doi.org/10.1140/epjc/s10052-016-4454-5} {\bibfield  {journal} {\bibinfo  {journal} {Eur. Phys. J. C}\ }\textbf {\bibinfo {volume} {76}},\ \bibinfo {pages} {652} (\bibinfo {year} {2016})}\BibitemShut {NoStop}%
\bibitem [{\citenamefont {Agostini}\ \emph {et~al.}(2024)\citenamefont {Agostini} \emph {et~al.}}]{GERDA2024}%
  \BibitemOpen
  \bibfield  {author} {\bibinfo {author} {\bibfnamefont {M.}~\bibnamefont {Agostini}} \emph {et~al.} (\bibinfo {collaboration} {GERDA Collaboration}),\ }\href {https://doi.org/10.1140/epjc/s10052-023-12280-6} {\bibfield  {journal} {\bibinfo  {journal} {Eur. Phys. J. C}\ }\textbf {\bibinfo {volume} {84}},\ \bibinfo {pages} {34} (\bibinfo {year} {2024})}\BibitemShut {NoStop}%
\bibitem [{\citenamefont {Krause}(1979)}]{electron_capture}%
  \BibitemOpen
  \bibfield  {author} {\bibinfo {author} {\bibfnamefont {M.~O.}\ \bibnamefont {Krause}},\ }\href {https://doi.org/10.1063/1.555594} {\bibfield  {journal} {\bibinfo  {journal} {J. Phys. Chem. Ref. Data}\ }\textbf {\bibinfo {volume} {8}},\ \bibinfo {pages} {307} (\bibinfo {year} {1979})}\BibitemShut {NoStop}%
\bibitem [{\citenamefont {Chen}(1991)}]{electron_capture2}%
  \BibitemOpen
  \bibfield  {author} {\bibinfo {author} {\bibfnamefont {M.~H.}\ \bibnamefont {Chen}},\ }\href {https://doi.org/10.1103/PhysRevA.44.239} {\bibfield  {journal} {\bibinfo  {journal} {Phys. Rev. A}\ }\textbf {\bibinfo {volume} {44}},\ \bibinfo {pages} {239} (\bibinfo {year} {1991})}\BibitemShut {NoStop}%
\bibitem [{\citenamefont {Crawford}\ \emph {et~al.}(2011)\citenamefont {Crawford}, \citenamefont {Cohen}, \citenamefont {Doherty},\ and\ \citenamefont {Atanacio}}]{electron_capture3}%
  \BibitemOpen
  \bibfield  {author} {\bibinfo {author} {\bibfnamefont {J.}~\bibnamefont {Crawford}}, \bibinfo {author} {\bibfnamefont {D.}~\bibnamefont {Cohen}}, \bibinfo {author} {\bibfnamefont {G.}~\bibnamefont {Doherty}},\ and\ \bibinfo {author} {\bibfnamefont {A.}~\bibnamefont {Atanacio}},\ }\href@noop {} {\emph {\bibinfo {title} {Calculated $K$-, $L$- and $M$-shell X-Ray line intensities for light line impact on selected targets from Z = 6 to 100}}},\ \bibinfo {type} {Tech. Rep.}\ \bibinfo {number} {ANSTO/E-774}\ (\bibinfo  {institution} {Australian Nuclear Science and Technology Organisation},\ \bibinfo {year} {2011})\BibitemShut {NoStop}%
\bibitem [{\citenamefont {Gagarin}\ and\ \citenamefont {Kovtun}(1981)}]{RAINE_code}%
  \BibitemOpen
  \bibfield  {author} {\bibinfo {author} {\bibfnamefont {S.~G.}\ \bibnamefont {Gagarin}}\ and\ \bibinfo {author} {\bibfnamefont {A.~P.}\ \bibnamefont {Kovtun}},\ }\href {https://doi.org/10.1007/BF00746365} {\bibfield  {journal} {\bibinfo  {journal} {J. Struct. Chem.}\ }\textbf {\bibinfo {volume} {21}},\ \bibinfo {pages} {1573} (\bibinfo {year} {1981})}\BibitemShut {NoStop}%
\bibitem [{\citenamefont {Band}\ and\ \citenamefont {Trzhaskovskaya}(1986)}]{RAINE}%
  \BibitemOpen
  \bibfield  {author} {\bibinfo {author} {\bibfnamefont {I.~M.}\ \bibnamefont {Band}}\ and\ \bibinfo {author} {\bibfnamefont {M.~B.}\ \bibnamefont {Trzhaskovskaya}},\ }\href {https://doi.org/10.1016/0092-640X(86)90027-6} {\bibfield  {journal} {\bibinfo  {journal} {At. Data Nucl. Data Tables}\ }\textbf {\bibinfo {volume} {35}},\ \bibinfo {pages} {1} (\bibinfo {year} {1986})}\BibitemShut {NoStop}%
\bibitem [{\citenamefont {Karpeshin}\ \emph {et~al.}(2012)\citenamefont {Karpeshin}, \citenamefont {Trzhaskovskaya},\ and\ \citenamefont {Kuz'minov}}]{Karpeshin}%
  \BibitemOpen
  \bibfield  {author} {\bibinfo {author} {\bibfnamefont {F.~F.}\ \bibnamefont {Karpeshin}}, \bibinfo {author} {\bibfnamefont {M.~B.}\ \bibnamefont {Trzhaskovskaya}},\ and\ \bibinfo {author} {\bibfnamefont {V.~V.}\ \bibnamefont {Kuz'minov}},\ }\href {https://doi.org/10.3103/S1062873812080187} {\bibfield  {journal} {\bibinfo  {journal} {Bull. Russ. Acad. Sci. Phys.}\ }\textbf {\bibinfo {volume} {76}},\ \bibinfo {pages} {884} (\bibinfo {year} {2012})}\BibitemShut {NoStop}%
\bibitem [{\citenamefont {Agnes}\ \emph {et~al.}(2018{\natexlab{a}})\citenamefont {Agnes} \emph {et~al.}}]{DarkSide:2018ppu}%
  \BibitemOpen
  \bibfield  {author} {\bibinfo {author} {\bibfnamefont {P.}~\bibnamefont {Agnes}} \emph {et~al.} (\bibinfo {collaboration} {DarkSide Collaboration}),\ }\href {https://doi.org/10.1103/PhysRevLett.121.111303} {\bibfield  {journal} {\bibinfo  {journal} {Phys. Rev. Lett.}\ }\textbf {\bibinfo {volume} {121}},\ \bibinfo {pages} {111303} (\bibinfo {year} {2018}{\natexlab{a}})}\BibitemShut {NoStop}%
\bibitem [{\citenamefont {Agnes}\ \emph {et~al.}(2018{\natexlab{b}})\citenamefont {Agnes} \emph {et~al.}}]{DarkSide:2018bpj}%
  \BibitemOpen
  \bibfield  {author} {\bibinfo {author} {\bibfnamefont {P.}~\bibnamefont {Agnes}} \emph {et~al.} (\bibinfo {collaboration} {DarkSide Collaboration}),\ }\href {https://doi.org/10.1103/PhysRevLett.121.081307} {\bibfield  {journal} {\bibinfo  {journal} {Phys. Rev. Lett.}\ }\textbf {\bibinfo {volume} {121}},\ \bibinfo {pages} {081307} (\bibinfo {year} {2018}{\natexlab{b}})}\BibitemShut {NoStop}%
\bibitem [{\citenamefont {Agnes}\ \emph {et~al.}(2023{\natexlab{a}})\citenamefont {Agnes} \emph {et~al.}}]{DarkSide:2022dhx}%
  \BibitemOpen
  \bibfield  {author} {\bibinfo {author} {\bibfnamefont {P.}~\bibnamefont {Agnes}} \emph {et~al.} (\bibinfo {collaboration} {DarkSide Collaboration}),\ }\href {https://doi.org/10.1103/PhysRevLett.130.101001} {\bibfield  {journal} {\bibinfo  {journal} {Phys. Rev. Lett.}\ }\textbf {\bibinfo {volume} {130}},\ \bibinfo {pages} {101001} (\bibinfo {year} {2023}{\natexlab{a}})}\BibitemShut {NoStop}%
\bibitem [{\citenamefont {Agnes}\ \emph {et~al.}(2023{\natexlab{b}})\citenamefont {Agnes} \emph {et~al.}}]{DarkSide:2022knj}%
  \BibitemOpen
  \bibfield  {author} {\bibinfo {author} {\bibfnamefont {P.}~\bibnamefont {Agnes}} \emph {et~al.} (\bibinfo {collaboration} {DarkSide Collaboration}),\ }\href {https://doi.org/10.1103/PhysRevLett.130.101002} {\bibfield  {journal} {\bibinfo  {journal} {Phys. Rev. Lett.}\ }\textbf {\bibinfo {volume} {130}},\ \bibinfo {pages} {101002} (\bibinfo {year} {2023}{\natexlab{b}})}\BibitemShut {NoStop}%
\bibitem [{\citenamefont {Agnes}\ \emph {et~al.}(2023{\natexlab{c}})\citenamefont {Agnes} \emph {et~al.}}]{DarkSide-50:2022qzh}%
  \BibitemOpen
  \bibfield  {author} {\bibinfo {author} {\bibfnamefont {P.}~\bibnamefont {Agnes}} \emph {et~al.} (\bibinfo {collaboration} {DarkSide-50 Collaboration}),\ }\href {https://doi.org/10.1103/PhysRevD.107.063001} {\bibfield  {journal} {\bibinfo  {journal} {Phys. Rev. D}\ }\textbf {\bibinfo {volume} {107}},\ \bibinfo {pages} {063001} (\bibinfo {year} {2023}{\natexlab{c}})}\BibitemShut {NoStop}%
\bibitem [{\citenamefont {Agnes}\ \emph {et~al.}(2023{\natexlab{d}})\citenamefont {Agnes} \emph {et~al.}}]{DarkSide-50:2023fcw}%
  \BibitemOpen
  \bibfield  {author} {\bibinfo {author} {\bibfnamefont {P.}~\bibnamefont {Agnes}} \emph {et~al.} (\bibinfo {collaboration} {DarkSide-50 Collaboration}),\ }\href {https://doi.org/10.1140/epjc/s10052-023-11410-4} {\bibfield  {journal} {\bibinfo  {journal} {Eur. Phys. J. C}\ }\textbf {\bibinfo {volume} {83}},\ \bibinfo {pages} {322} (\bibinfo {year} {2023}{\natexlab{d}})}\BibitemShut {NoStop}%
\bibitem [{\citenamefont {Acerbi}\ \emph {et~al.}(2025)\citenamefont {Acerbi} \emph {et~al.}}]{2025sensitivitylowmasswimpsimproved}%
  \BibitemOpen
  \bibfield  {author} {\bibinfo {author} {\bibfnamefont {F.}~\bibnamefont {Acerbi}} \emph {et~al.} (\bibinfo {collaboration} {DarkSide Collaboration})} (\bibinfo {year} {2025}),\ \bibinfo {note} {arXiv:2511.13629}\BibitemShut {NoStop}%
\bibitem [{\citenamefont {Bellini}\ \emph {et~al.}(2013)\citenamefont {Bellini} \emph {et~al.}}]{underground}%
  \BibitemOpen
  \bibfield  {author} {\bibinfo {author} {\bibfnamefont {G.}~\bibnamefont {Bellini}} \emph {et~al.} (\bibinfo {collaboration} {Borexino Collaboration}),\ }\href {https://doi.org/10.1088/1475-7516/2013/08/049} {\bibfield  {journal} {\bibinfo  {journal} {JCAP}\ }\textbf {\bibinfo {volume} {2013}}\bibinfo  {number} { (8)},\ \bibinfo {pages} {049}}\BibitemShut {NoStop}%
\bibitem [{\citenamefont {Agnes}\ \emph {et~al.}(2015)\citenamefont {Agnes} \emph {et~al.}}]{TPC}%
  \BibitemOpen
\bibfield  {number} {  }\bibfield  {author} {\bibinfo {author} {\bibfnamefont {P.}~\bibnamefont {Agnes}} \emph {et~al.} (\bibinfo {collaboration} {DarkSide Collaboration}),\ }\href {https://doi.org/10.1016/j.physletb.2015.03.012} {\bibfield  {journal} {\bibinfo  {journal} {Phys. Lett. B}\ }\textbf {\bibinfo {volume} {743}},\ \bibinfo {pages} {456} (\bibinfo {year} {2015})}\BibitemShut {NoStop}%
\bibitem [{\citenamefont {Begemann}\ \emph {et~al.}(1976)\citenamefont {Begemann}, \citenamefont {Weber},\ and\ \citenamefont {Hintenberger}}]{primordial40Ar}%
  \BibitemOpen
  \bibfield  {author} {\bibinfo {author} {\bibfnamefont {F.}~\bibnamefont {Begemann}}, \bibinfo {author} {\bibfnamefont {H.}~\bibnamefont {Weber}},\ and\ \bibinfo {author} {\bibfnamefont {H.}~\bibnamefont {Hintenberger}},\ }\href {https://doi.org/10.1086/182041} {\bibfield  {journal} {\bibinfo  {journal} {Astrophys. J.}\ }\textbf {\bibinfo {volume} {203}},\ \bibinfo {pages} {L155} (\bibinfo {year} {1976})}\BibitemShut {NoStop}%
\bibitem [{\citenamefont {Atreya}\ \emph {et~al.}(2013)\citenamefont {Atreya} \emph {et~al.}}]{primordial_Mars}%
  \BibitemOpen
  \bibfield  {author} {\bibinfo {author} {\bibfnamefont {F.~H.~B.}\ \bibnamefont {Atreya}, \bibfnamefont {S.~K. Trainer M.~G.}} \emph {et~al.},\ }\href {https://doi.org/10.1002/2013GL057763} {\bibfield  {journal} {\bibinfo  {journal} {Geophys. Res. Lett.}\ }\textbf {\bibinfo {volume} {40}},\ \bibinfo {pages} {5605} (\bibinfo {year} {2013})}\BibitemShut {NoStop}%
\bibitem [{\citenamefont {Marty}(2012)}]{Marty_2012}%
  \BibitemOpen
  \bibfield  {author} {\bibinfo {author} {\bibfnamefont {B.}~\bibnamefont {Marty}},\ }\href {https://doi.org/10.1016/j.epsl.2011.10.040} {\bibfield  {journal} {\bibinfo  {journal} {Earth Planet. Sci. Lett.}\ }\textbf {\bibinfo {volume} {313-314}},\ \bibinfo {pages} {56} (\bibinfo {year} {2012})}\BibitemShut {NoStop}%
\bibitem [{\citenamefont {Adhikari}\ \emph {et~al.}(2025)\citenamefont {Adhikari} \emph {et~al.}}]{DEAP_Ar39}%
  \BibitemOpen
  \bibfield  {author} {\bibinfo {author} {\bibfnamefont {P.}~\bibnamefont {Adhikari}} \emph {et~al.} (\bibinfo {collaboration} {DEAP Collaboration}),\ }\href {https://doi.org/10.1140/epjc/s10052-025-14289-5} {\bibfield  {journal} {\bibinfo  {journal} {Eur. Phys. J. C}\ }\textbf {\bibinfo {volume} {85}},\ \bibinfo {pages} {728} (\bibinfo {year} {2025})}\BibitemShut {NoStop}%
\bibitem [{\citenamefont {Sarda}\ \emph {et~al.}(1985)\citenamefont {Sarda}, \citenamefont {Staudacher},\ and\ \citenamefont {Allègre}}]{UAr_Allegre}%
  \BibitemOpen
  \bibfield  {author} {\bibinfo {author} {\bibfnamefont {P.}~\bibnamefont {Sarda}}, \bibinfo {author} {\bibfnamefont {T.}~\bibnamefont {Staudacher}},\ and\ \bibinfo {author} {\bibfnamefont {C.~J.}\ \bibnamefont {Allègre}},\ }\href {https://doi.org/10.1016/0012-821X(85)90058-5} {\bibfield  {journal} {\bibinfo  {journal} {Earth Planet. Sci. Lett.}\ }\textbf {\bibinfo {volume} {72}},\ \bibinfo {pages} {357} (\bibinfo {year} {1985})}\BibitemShut {NoStop}%
\bibitem [{\citenamefont {Hu}\ and\ \citenamefont {Moynier}(2025)}]{ICP-MS}%
  \BibitemOpen
  \bibfield  {author} {\bibinfo {author} {\bibfnamefont {Y.}~\bibnamefont {Hu}}\ and\ \bibinfo {author} {\bibfnamefont {F.}~\bibnamefont {Moynier}},\ }in\ \href {https://doi.org/10.1016/B978-0-323-99762-1.00063-2} {\emph {\bibinfo {booktitle} {Treatise on Geochemistry}}},\ \bibinfo {editor} {edited by\ \bibinfo {editor} {\bibfnamefont {A.}~\bibnamefont {Anbar}}\ and\ \bibinfo {editor} {\bibfnamefont {D.}~\bibnamefont {Weis}}}\ (\bibinfo  {publisher} {Elsevier},\ \bibinfo {address} {Oxford},\ \bibinfo {year} {2025})\ \bibinfo {edition} {3rd}\ ed.,\ pp.\ \bibinfo {pages} {497--545}\BibitemShut {NoStop}%
\bibitem [{\citenamefont {Santorelli}\ \emph {et~al.}(2023)\citenamefont {Santorelli}, \citenamefont {Olmedo}, \citenamefont {Vilda}, \citenamefont {Diaz}, \citenamefont {Pesudo},\ and\ \citenamefont {Romero}}]{Santorelli:2023vzj}%
  \BibitemOpen
  \bibfield  {author} {\bibinfo {author} {\bibfnamefont {R.}~\bibnamefont {Santorelli}}, \bibinfo {author} {\bibfnamefont {A.~I.~B.}\ \bibnamefont {Olmedo}}, \bibinfo {author} {\bibfnamefont {E.~C.}\ \bibnamefont {Vilda}}, \bibinfo {author} {\bibfnamefont {M.~F.}\ \bibnamefont {Diaz}}, \bibinfo {author} {\bibfnamefont {V.}~\bibnamefont {Pesudo}},\ and\ \bibinfo {author} {\bibfnamefont {L.}~\bibnamefont {Romero}},\ }\href {https://doi.org/10.1063/5.0161118} {\bibfield  {journal} {\bibinfo  {journal} {AIP Conf. Proc.}\ }\textbf {\bibinfo {volume} {2908}},\ \bibinfo {pages} {080003} (\bibinfo {year} {2023})}\BibitemShut {NoStop}%
\bibitem [{\citenamefont {Agnes}\ \emph {et~al.}(2021{\natexlab{a}})\citenamefont {Agnes} \emph {et~al.}}]{response_calibration}%
  \BibitemOpen
  \bibfield  {author} {\bibinfo {author} {\bibfnamefont {P.}~\bibnamefont {Agnes}} \emph {et~al.} (\bibinfo {collaboration} {DarkSide Collaboration}),\ }\href {https://doi.org/10.1103/PhysRevD.104.082005} {\bibfield  {journal} {\bibinfo  {journal} {Phys. Rev. D}\ }\textbf {\bibinfo {volume} {104}},\ \bibinfo {pages} {082005} (\bibinfo {year} {2021}{\natexlab{a}})}\BibitemShut {NoStop}%
\bibitem [{\citenamefont {Cranmer}\ \emph {et~al.}(2012)\citenamefont {Cranmer}, \citenamefont {Lewis} \emph {et~al.}}]{RooFit}%
  \BibitemOpen
  \bibfield  {author} {\bibinfo {author} {\bibfnamefont {K.}~\bibnamefont {Cranmer}}, \bibinfo {author} {\bibfnamefont {G.}~\bibnamefont {Lewis}}, \emph {et~al.},\ }\href {https://doi.org/10.17181/CERN-OPEN-2012-016} {\emph {\bibinfo {title} {HistFactory: A tool for creating statistical models for use with RooFit and RooStats}}},\ \bibinfo {type} {Tech. Rep.}\ (\bibinfo {year} {2012})\BibitemShut {NoStop}%
\bibitem [{\citenamefont {Cowan}\ \emph {et~al.}(2011)\citenamefont {Cowan}, \citenamefont {Cranmer}, \citenamefont {Gross},\ and\ \citenamefont {Vitells}}]{Cowan_2011}%
  \BibitemOpen
  \bibfield  {author} {\bibinfo {author} {\bibfnamefont {G.}~\bibnamefont {Cowan}}, \bibinfo {author} {\bibfnamefont {K.}~\bibnamefont {Cranmer}}, \bibinfo {author} {\bibfnamefont {E.}~\bibnamefont {Gross}},\ and\ \bibinfo {author} {\bibfnamefont {O.}~\bibnamefont {Vitells}},\ }\href {https://doi.org/10.1140/epjc/s10052-011-1554-0} {\bibfield  {journal} {\bibinfo  {journal} {Eur. Phys. J. C}\ }\textbf {\bibinfo {volume} {71}},\ \bibinfo {pages} {1554} (\bibinfo {year} {2011})}\BibitemShut {NoStop}%
\bibitem [{\citenamefont {Aalseth}\ \emph {et~al.}(2018)\citenamefont {Aalseth} \emph {et~al.}}]{DS-20k}%
  \BibitemOpen
  \bibfield  {author} {\bibinfo {author} {\bibfnamefont {C.~E.}\ \bibnamefont {Aalseth}} \emph {et~al.} (\bibinfo {collaboration} {DarkSide-20k Collaboration}),\ }\href {https://doi.org/10.1140/epjp/i2018-11973-4} {\bibfield  {journal} {\bibinfo  {journal} {Eur. Phys. J. Plus}\ }\textbf {\bibinfo {volume} {133}},\ \bibinfo {pages} {131} (\bibinfo {year} {2018})}\BibitemShut {NoStop}%
\bibitem [{\citenamefont {Agnes}\ \emph {et~al.}(2021{\natexlab{b}})\citenamefont {Agnes} \emph {et~al.}}]{ARIA}%
  \BibitemOpen
  \bibfield  {author} {\bibinfo {author} {\bibfnamefont {P.}~\bibnamefont {Agnes}} \emph {et~al.} (\bibinfo {collaboration} {DarkSide-20k Collaboration}),\ }\href {https://doi.org/10.1140/epjc/s10052-021-09121-9} {\bibfield  {journal} {\bibinfo  {journal} {Eur. Phys. J. C}\ }\textbf {\bibinfo {volume} {81}},\ \bibinfo {pages} {359} (\bibinfo {year} {2021}{\natexlab{b}})}\BibitemShut {NoStop}%
\bibitem [{\citenamefont {Agnes}\ \emph {et~al.}(2025)\citenamefont {Agnes} \emph {et~al.}}]{spurious_electrons}%
  \BibitemOpen
  \bibfield  {author} {\bibinfo {author} {\bibfnamefont {P.}~\bibnamefont {Agnes}} \emph {et~al.} (\bibinfo {collaboration} {DarkSide-50 Collaboration})} (\bibinfo {year} {2025}),\ \bibinfo {note} {arXiv:2507.23003}\BibitemShut {NoStop}%
\bibitem [{\citenamefont {Acerbi}\ \emph {et~al.}(2024)\citenamefont {Acerbi} \emph {et~al.}}]{DS20k_sensitivity}%
  \BibitemOpen
  \bibfield  {author} {\bibinfo {author} {\bibfnamefont {F.}~\bibnamefont {Acerbi}} \emph {et~al.} (\bibinfo {collaboration} {DarkSide-20k Collaboration}),\ }\href {https://doi.org/10.1038/s42005-024-01896-z} {\bibfield  {journal} {\bibinfo  {journal} {Commun. Phys.}\ }\textbf {\bibinfo {volume} {7}},\ \bibinfo {pages} {260} (\bibinfo {year} {2024})}\BibitemShut {NoStop}%
\end{thebibliography}
%

\end{document}